\documentstyle[12pt]{article}
\newcommand{\gsim}{\mbox{\raisebox{-.3em}{$\stackrel{>}{\sim}$}}}
\newcommand{\lsim}{\mbox{\raisebox{-.3em}{$\stackrel{<}{\sim}$}}}

\newcommand{\beq}{\begin{equation}}
\newcommand{\eeq}{\end{equation}}
\newcommand{\beqa}{\begin{eqnarray}}
\newcommand{\eeqa}{\end{eqnarray}}
\newcommand{\bpc}{\begin{picture}}
\newcommand{\epc}{\end{picture}}
\newcommand{\bfg}{\begin{figure}}
\newcommand{\efg}{\end{figure}}
\newcommand{\bcent}{\begin{center}}
\newcommand{\ecent}{\end{center}}

\newcommand{\reflef}{(\ref}
\newcommand{\nnb}{\nonumber}

\begin{document}    
\renewcommand{\topfraction}{0.95}
\renewcommand{\bottomfraction}{0.95}
\renewcommand{\textfraction}{0.05}
\renewcommand{\dbltopfraction}{0.95}
\renewcommand{\textfraction}{0.05}
\input psfig.tex    
\begin{center}      
{\large\bf The nuclear interaction at Oklo 2 billion years ago\footnote{A
preliminary report is found in Ref.~[1].}}\\[0.75em]
Yasunori Fujii\\    
Nihon Fukushi University, Handa, Aichi, 475-0012  Japan \\
and \\              
Institute of Cosmic Ray Research (ICRR), University of Tokyo,\\
Tanashi, Tokyo, 188-8502 Japan\\[0.5em]
Akira Iwamoto, Tokio Fukahori, Toshihiko Ohnuki, Masayuki Nakagawa\\
Japan Atomic Energy Research Institute (JAERI)\\
Tokai-mura, Naka-gun, Ibaraki, 319-1195 Japan \\[0.5em]
 Hiroshi Hidaka\\   
Department of Earth and Planetary Systems Science, Hiroshima University\\
Higashi-Hiroshima, Hiroshima, 739-8526 Japan\\[0.5em]
Yasuji Oura\\       
Department of Chemistry, Tokyo Metropolitan University,\\
Hachioji, Tokyo,    
192-0397 Japan\\[0.5em]
Peter M\"{o}ller\\  
Theoretical Division, Los Alamos National Laboratory, New Mexico 87545, USA\\
\end{center}        
\bigskip            
\bcent              
{\large\bf Abstract}\\
\bigskip            
\begin{minipage}{13cm}
We re-examine the effort to constrain the
time-variability of the  coupling constants of the fundamental
interactions by  studying the
anomalous  isotopic abundance of Sm observed at the remnants of the
natural reactors which were in operation at Oklo about 2 billion years
ago, in terms of a possible deviation of the resonance energy from
the value observed today.  We rely on  new samples that were carefully collected  to minimize
natural contamination and also on a careful temperature estimate of the
 reactors.  We obtain the upper bound  $(-0.2\pm 0.8)\times 10^{-17}$~${\rm
y}^{-1}$  on the fractional rate of change of the electromagnetic as well as
the strong interaction coupling constants.  Our result
basically agrees with and even suggests some improvement of the result
due recently to Damour and Dyson.   Strictly speaking, however, we
 find another range of the resonance energy shift indicating a nonzero
time variation of the constants.  We find a rather strong but still tentative 
indication that this non-null range can be ruled 
out by including  the Gd data, for which it is essential to take the effect of
contamination into account.

\end{minipage}      
 
\ecent              
 
\newpage            
\section{Introduction}

Few aspects of physics can be understood without  recourse to
fundamental ``constants,'' such as the speed of light $c$,
Planck's constant $h$, Newton's
constant $G$, the fine-structure constant $\alpha$, and its QCD
counterpart $\alpha_{\rm QCD}$, or the pion-nucleon coupling constant
squared $\alpha_{\rm s}=g_{\pi N}^2/4\pi$ in most of realistic analyses.
To this list one may also add the  masses
of certain fundamental particles, like the quarks and the leptons, or some of
the vacuum expectation values, and perhaps the cosmological constant
$\Lambda$.  The traditional assumption that these parameters of physics theories
are constant        
 was challenged for the first time by Dirac,
who suggested that $G$ might vary as $t^{-1}$  with $t$ the cosmic
time [2].  This paper inspired to
many probes  of the possible time variability of various fundamental constants.

However, no positive evidence for any time-dependence has been
discovered so far, nor does Dirac's  original argument seem fully convincing.
Nevertheless, the very idea that at least some of the fundamental
constants may not be truly constant still attracts serious attention
partly because this appears to be a rather general outcome of the
recent efforts toward unifying particle physics and gravity.

Quite remarkable is the finding that the upper bounds obtained so far
for $\alpha$ and $\alpha_{\rm s}$
are many orders of magnitude  below  the value $\sim 10^{-10}$~${\rm y}^{-1}$
expected naturally in
terms of the present age of the universe $t_0 \sim 10^{10}$~${\rm y}$
[3--9].             
The same tendency is indicated also for $G$ [10].
This  is expected to provide an important clue to the nature of a model of unification.

The most stringent constraint ever reported is
due to Shlyakhter [6,7] who exploited the exceptionally
sensitive dependence of the cross section of the reaction
\beq                
{\rm n} + {\rm \mbox{$^{149}$}Sm} \rightarrow  {\rm \mbox{$^{150}$}Sm}+ \gamma,
\label{okp1_1}      
\eeq                
on the energy of a resonance lying  as low as
$E_{\rm r}=97.3$~meV above threshold, which
corresponds to the temperature $856$$^{\circ}$C and is much
smaller than any of the mass scales of the strong or the
electromagnetic interaction.

It was observed that the anomalous abundance of $^{149}{\rm Sm}$
observed at Oklo [11,12] can be understood rather well in terms of the
reaction \reflef{okp1_1}) with the fundamental-constant values {\em observed today}.  An analysis of the uncertainties in the Oklo observations  provides an upper bound on the possible deviation of the neutron-capture cross section $\sigma_{149}$ associated with the reaction
\reflef{okp1_1}) occurring 2 billions years ago from its present value.
Shlyakhter concluded that    
$|\Delta\sigma_{149} /\sigma_{149}|\lsim 10\%$, which  implies
$|\Delta E_{\rm r}|\lsim 20$~meV.  Furthermore he derived that
$|\Delta\alpha_{\rm s} /\alpha_{\rm s}|\lsim 5\times 10^{-10}$~${\rm y}^{-1}$, or $|\dot{\alpha}_s/\alpha_{\rm s}| \lsim 2.5 \times 10^{-19}$~${\rm y}^{-1}$,
which represent upper bounds on the time-variability of the
fundamental constant that are  several orders of magnitude more
stringent than any other estimate of these bounds [3--5,8,9].

However, this substantial improvement of the upper-bound estimate
does not seem fully appreciated,
because complete details of the analysis have never been  published. Another
reason for the limited acceptance may be that the derivation steps are
much less direct than in estimates based on quasi-stellar-object
spectra [8] or on clock standards [9].

More recently Damour and Dyson [13] repeated the analysis 
on the same process \reflef{okp1_1}) in greater detail, reaching a
somewhat more conservative estimate, $-120 {\rm meV}< \Delta E_{\rm r}
< 90 {\rm meV}.$  They also showed that the result can be interpreted
as an upper bound  $|\dot{\alpha}/\alpha |\lsim 6\times 10^{-17}{\rm 
y}^{-1}$, through the time variation of the Coulomb energy which can
be analyzed on a much more solid theoretical basis than for the strong-interaction coupling constant $\alpha_{\rm s}$.

Having learned, however, about the availability of the data taken even more recently, we decided to pursue the subject further.  The 
data are samples obtained from deep underground  with a geologist's expertise
so as to limit  potential outside contamination  as much as possible.
We also try to use 
the data from other isotopes to confirm the result obtained from Sm.  Possible check of the effect of outside contamination is also planned.

We find two ranges for the allowed values of $\Delta E_{\rm r}$, one
that covers the null result and another apparently indicating a
nonzero effect. We try to see if we can exclude one of them by
 the aid of the Gd data which is admittedly subject to more
uncertainties than the Sm data, due to much smaller abundance because
of the more significant absorption effect.  However, by taking
advantage of the presence of the two low-lying isotope resonances, we
are able to suggest that the range around  $\Delta E_{\rm r}=0$ is
favored if we accept certain plausibility arguments.  A narrower
constraint on the reactor temperatures obtained recently provides an
added support to this conclusion.

We also show that the theoretical analysis designed originally to be
applied to constrain the variability of the electromagnetic coupling
constant [13] can be applied to the strong interaction as well, thanks
to the exceptionally small value of $E_{\rm r}$, requiring a near cancellation between the effects of the two interactions.

 We introduce basic equations that are used in our analyses: in
Section 2 those related to neutron absorption and in Section 3 those
related to nuclide transformations along isotope chains.  
New measurements on five relatively recent samples are
then presented and analyzed in Section 4, providing a better upper bound on
$\Delta E_{\rm r}$. 
Section 5 discusses the relevance of the results to the possible time-variation
of  the electromagnetic-interaction and the strong-interaction coupling constants.  Section 6 gives discussions summary of the results obtained in our investigation.  Appendix A provides some details of the numerical
calculation of the thermal average of the resonance cross sections.
In Appendix B we show  some details of the fluence determination adopted in this investigation.

 
\section{Neutron absorption cross sections}

The results obtained in this
investigation are based on an
analysis of the neutron-absorption cross section in the reaction \reflef{okp1_1}) and
the corresponding reactions on Gd. The reactions are dominated by
 a resonance, which we describe  in the standard Breit-Wigner form,
\beq                
\sigma_{\rm r} =\frac{g_0 \pi \hbar^2}{2m E}\frac{\Gamma_{\rm n}
\Gamma_{\gamma}}{(E-E_{\rm r})^2
 + \Gamma_{\rm tot}^2 /4},
\label{okp2_1}      
\eeq                
where $g_0= (2J+1)/[(2s+1)(2I+1)]$ is the statistical factor,
$s$ the spin of the neutron, $I$ the spin of the target, $J$ the spin of
the compound nucleus, and
$\Gamma_{\rm tot} =\Gamma_{\rm n} + \Gamma_{\gamma}$
the total width     
in terms of the neutron
and the photon widths, respectively.  Also, $\Gamma_{\rm n}
=(2k/K)(D/\pi)$, where $k^2 = 2mE/\hbar^2$,  $K$  the wave number
inside the target nucleus,  and  $D$  the level spacing of the compound
levels.   The values of these  parameters
are given in Table \ref{table1}  for the lowest-lying resonance in
\begin{table}[t]    
\bcent              
\begin{tabular}{||clc |c  r r r c||} \hline
& & & & &  &&\\[-0.07in]
& Channel& & &$^{149}{\rm Sm}$ & $^{155}{\rm Gd}$ & $^{157}{\rm Gd}$& \\[0.08in]
\hline              
& & & & &&& \\[-0.07in]
&Resonance energy $E_{\rm r}$ (meV)&  & & 97.3 & 26.8 & 31.4& \\
&Neutron width $\Gamma_{\rm n}$ (meV)& & &0.533 & 0.104 & 0.470& \\
&Gamma width $\Gamma_{\gamma}$ (meV)& && 60.5 & 108 & 106& \\
&Statistical factor $g_0$& & &9/16 & 5/8 & 5/8 &\\[0.08in]
\hline              
\end{tabular}       
\ecent              
\caption[ctab1]{\baselineskip=12pt\small
Parameters of the  lowest-lying resonances used in the present
analysis. The tabulated values of $E_{\rm r}$  are the present
resonance-position values $E_{\rm r0}$ [14].
The listed values of $\Gamma_{\rm n}$ are estimated at  $E_{\rm r}$.}     
\label{table1}      
\end{table}         
the reactions on $^{149}{\rm Sm}$, $^{155\!}{\rm Gd}$ and $^{157}{\rm Gd}$.
 
We now derive       
how the isotopic abundances observed today
depend on $E_{\rm r}$, since we need to consider the possibility that
$E_{\rm r}$  were different from its present value 2 billion years ago.
Other parameters may be  assumed constant, since only $E_{\rm r}$ will
affect the result in any significant manner.  Note that we would
normally expect abundance changes  not much larger than  $\Delta\alpha
/\alpha$, unless they are amplified by the resonance mechanism
and by the extremely small value of  $E_{\rm r}$ relative to
any of the mass scales of nuclear physics.

We average \reflef{okp2_1})  with respect to the thermal neutron flux,
\beq                
\Phi(E,T) =v(E) \rho(E,T),\quad \mbox{with}\quad v=\sqrt{2E/m},
\label{okp2_4}      
\eeq                
where               
\beq                
\rho(E,T)=2\pi^{-1/2}T^{-3/2} e^{-E/T}\sqrt{E}
\label{okp2_5}      
\eeq                
is the normalized Maxwell-Boltzmann distribution.

Normally the thermally averaged cross section is defined by
\beq                
\sigma(T)= [{\cal D}(T)]^{-1} \int \sigma (E)\Phi(E,T)dE,
\label{okp2_6}      
\eeq                
where               
\beq                
{\cal D}(T)= \int \Phi(E,T)dE.
\label{okp2_7}      
\eeq

However, in most of the analyses of the Oklo phenomenon,  it is
customary [15] to replace  the denominator ${\cal D}(T)$  by
\beq                
\hat{\cal D}= v_0 \int \rho(E, T) dE = v_0,
\label{okp2_8}      
\eeq                
for a convenient choice of the velocity $v_0 =2200$~m/s corresponding
to the temperature $T_0 =  (m/2k)v_0^2
= 20.4^{\circ}$C, hence defining
an {\it effective cross section}
\beq                
\hat{\sigma}(T)=\hat{\cal D}^{-1}\int \sigma_{\rm r} (E)\Phi(E,T)dE
=\sqrt{\frac{4}{\pi}\frac{T}{T_0}} \sigma (T).
\label{okp2_9}      
\eeq                
 
The advantage of this non-standard definition is
that, unlike ${\cal D}(T)$,
$\hat{\cal D}$ is a {\em constant}, independent of $T$.
We note that the neutron flux at Oklo 2 billion years ago is determined  based on the observed
abundances of $^{143,145}{\rm Nd}$ and $^{147}{\rm Sm}$, as will be
shown in more detail in Appendix B.  The corresponding neutron-absorption
cross sections are known to obey a $1/v$ behavior to a good
approximation, implying that the {\em effective} cross sections of these
processes are nearly
$T$-independent, hence simplifying the practical analysis considerably.

In the following analysis, the cross sections always occur multiplied by
the thermally averaged flux $\phi(T)$. It is obviously convenient to
introduce an {\it effective flux} $\hat{\phi}$ in such a way that the
product with  $\hat{\sigma}$ remains the same as the original product:
\beq                
\sigma\phi = \hat{\sigma}\hat{\phi}.
\label{okp2_10}     
\eeq                
Using the last expression of \reflef{okp2_9}), we find
\beq                
\hat{\phi}= \sqrt{\frac{\pi}{4}\frac{T_0}{T}}\phi.
\label{okp2_11}     
\eeq

We evaluate the integral in \reflef{okp2_9}) by  numerical integration.
Details             
of the calculation  
are found in        
Appendix A.         
The results for     
\begin{figure}[tb]  
\centerline{\psfig{figure=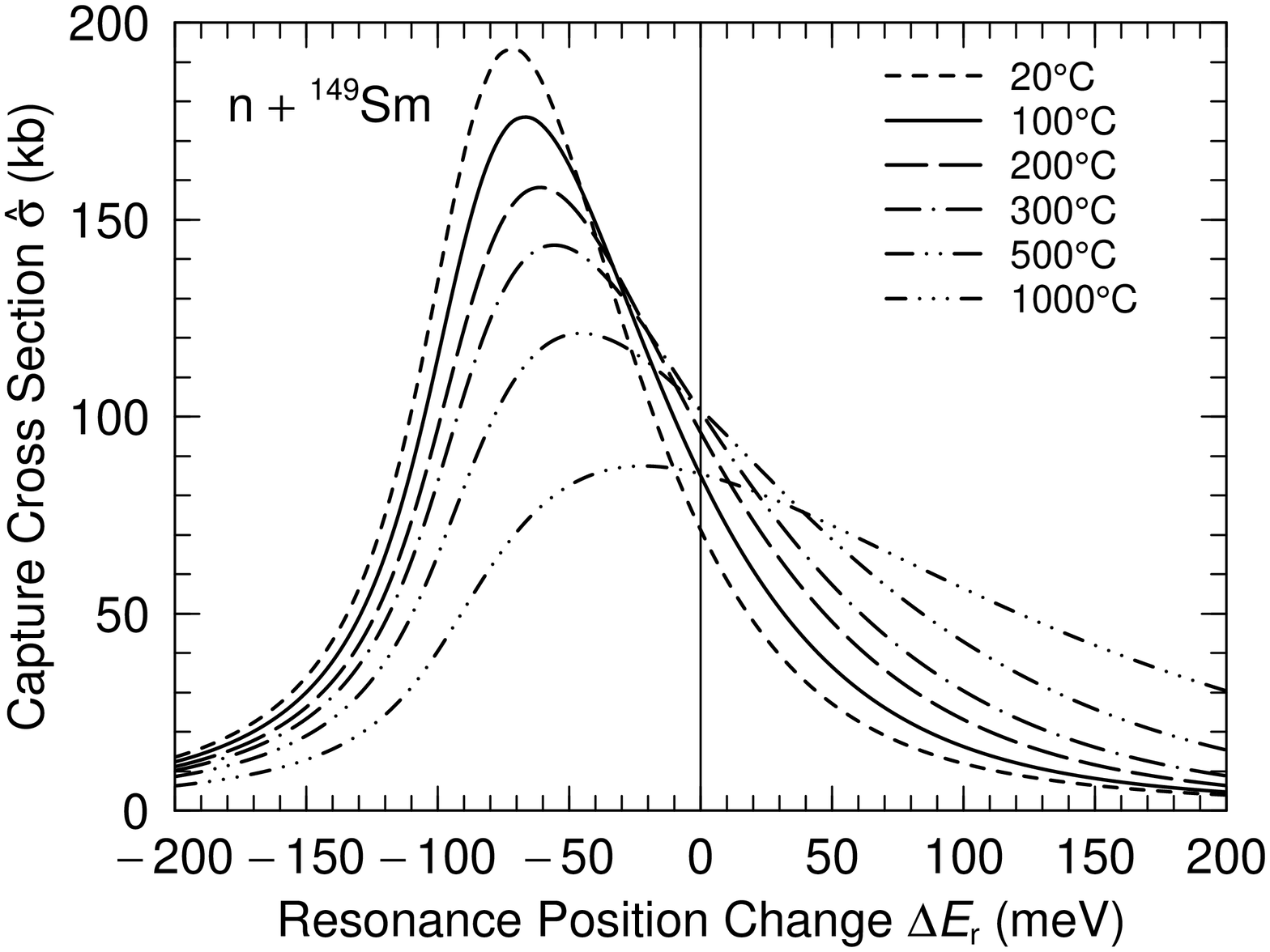,width=6.3in}}
\caption[cfig1]{\baselineskip=12pt\small
Calculated thermally
averaged cross sections $\hat{\sigma}_{149}$ for ${\rm n}+
^{149\!}{\rm Sm}\rightarrow ^{150\!}{\rm Sm} + \gamma$ as  functions
of the resonance position change $\Delta E_{\rm r}$ and the temperature $T$.  }
\label{fig1}        
\end{figure} 
 reactions on $^{149}$Sm, $^{155}$Gd, and $^{157}$Gd
\begin{figure}[t]   
\centerline{\psfig{figure=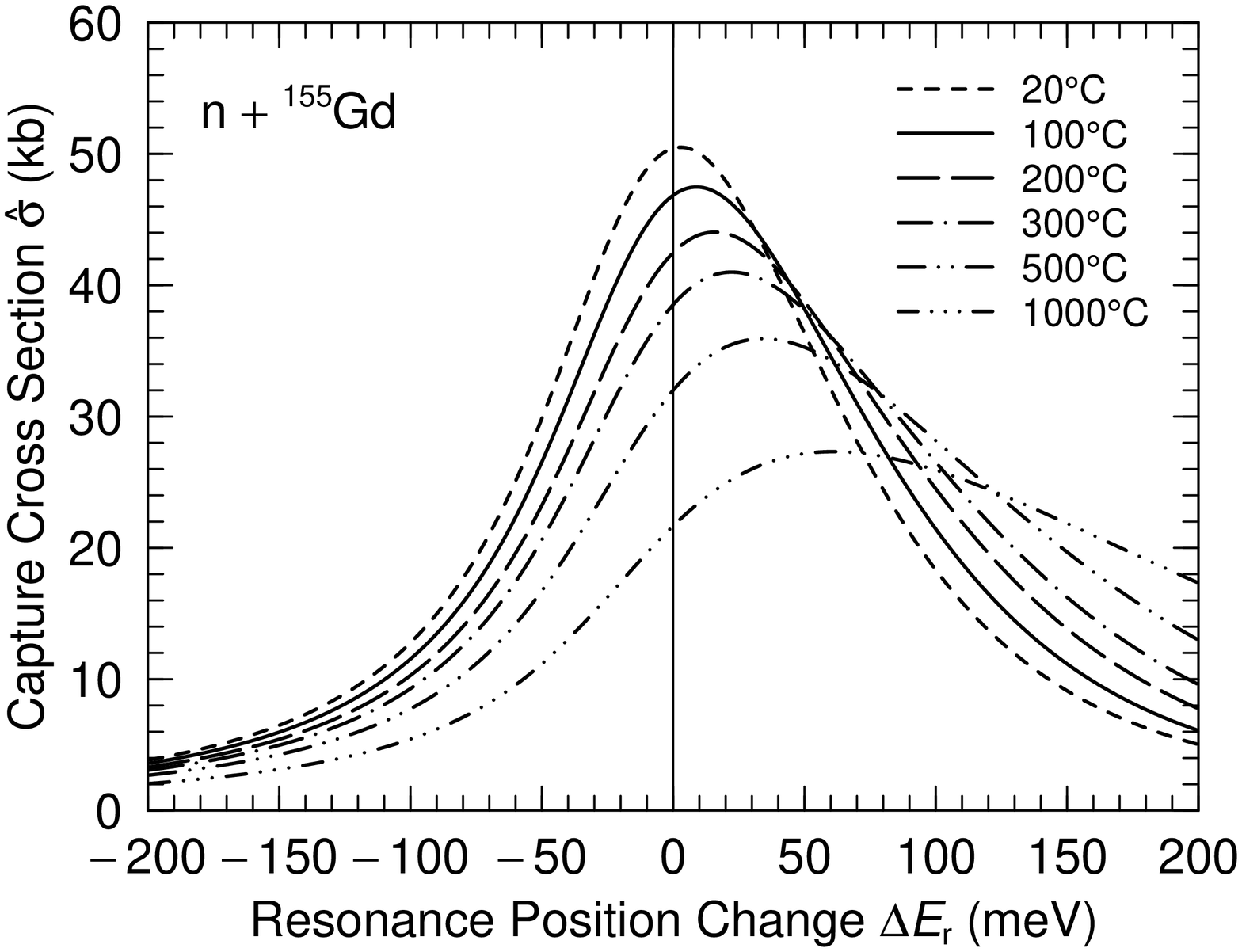,width=6.3in}}
\caption[cfig2]{\baselineskip=12pt\small
Calculated thermally averaged cross sections $\hat{ \sigma}_{155}$ for
 ${\rm n}+ ^{155\!}{\rm Gd}\rightarrow ^{156\!}{\rm Gd} + \gamma$
as  functions       
of the resonance position change $\Delta E_{\rm r}$ and the temperature $T$.  }
\label{fig2}        
\end{figure}
are shown  in Figs.~\ref{fig1}--\ref{fig3}, respectively.
\begin{figure}[t]   
\centerline{\psfig{figure=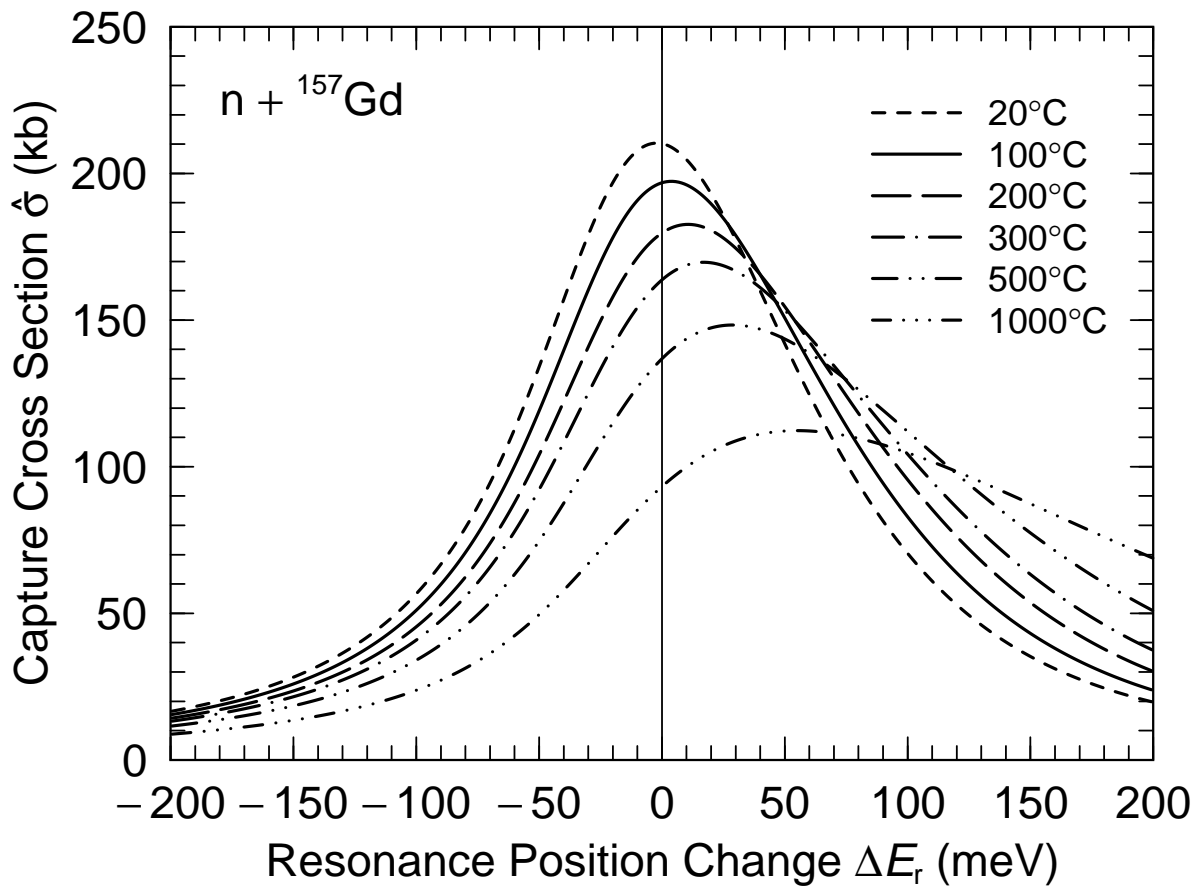,width=6.3in}}
\caption[cfig3]{\baselineskip=12pt\small
Calculated thermally averaged cross sections $\hat{ \sigma}_{157}$
for ${\rm n}+ ^{157\!}{\rm Gd}\rightarrow ^{158\!}{\rm Gd} + \gamma$
as  functions       
of the resonance position change $\Delta E_{\rm r}$ and the temperature $T$.  }
\label{fig3}        
\end{figure} 
The figures illustrate the strong dependence of the thermally-averaged cross section  on
the resonance-position change $\Delta E_{\rm r}$  for various reactor temperatures $T$.
 
We have calculated numerically the effect of thermal fluctuations of
the atoms in the neutron absorber   on the calculated cross
sections. The magnitude of this so-called Doppler effect on the
calculated $\hat{\sigma}(T)$ is less than 1\%
even at the highest temperature investigated, 1000$^{\circ}$C, and much smaller
for lower temperatures. Therefore this effect can be safely ignored,
and the formalism below is for simplicity developed without including Doppler broadening.

\section{Relations for isotopic compositions}
 
The isotope ratios present in Oklo today are superpositions of contributions
from fission, subsequent neutron capture, and of the original natural
abundances. The interpretation of the observed abundances is simplified when in
special cases some processes do not contribute. In a few of those cases,
where the various contributions to the now observed isotope ratios are most
easily disentangled, we may
draw stringent conclusions about the variation in time of a few
low-lying neutron-resonance positions. We consider here the cases of neutron
capture on $^{149}$Sm, $^{155}$Gd, and $^{157}$Gd.

\subsection{Isotopic composition of Sm}

We can make a few observations that simplify
the differential equations obtained for the time evolution of the number of
atoms per unit volume $N_A(t)$ of
the samarium isotopes $^A{\rm Sm}$  near $^{149}$Sm that are
connected by nuclear transformations.
The {\it spontaneous-fission} half-lives of $^{235}$U and $^{238}$U
exceed  $10^{16}$ y. Therefore fission products are generated in
appreciable amounts only by neutron-induced fission
during the period of operation of the natural reactors at Oklo.
Some isotopes are shielded by  stable nuclides and cannot be reached
in the $\beta$-decay of the unstable nuclei created during the fission process.
In addition, it is possible to neglect some neutron-capture reactions because of
their               
low cross sections,  2.413 barn~[16], for example, for
capture on $^{148}$Sm.

Thus, we obtain the following differential equations:
\beqa               
\frac{dN_{147}(t)}{dt}&=&-\hat{\sigma}_{147}\hat{\phi}N_{147}(t) +
N^0_{235}\exp (-\hat{\sigma}_{\rm a}\hat{\phi}t) \hat{\sigma}_{\rm f235}\hat{\phi}Y_{147}, \nnb\\
\frac{dN_{148}(t)}{dt}&=&\hat{\sigma}_{147}\hat{\phi}N_{147}(t), \nnb\\
\frac{dN_{149}(t)}{dt}&=&-\hat{\sigma}_{149}\hat{\phi}N_{149}(t) +
N^0_{235}\exp (-\hat{\sigma}_{\rm a}\hat{\phi}t) \hat{\sigma}_{\rm f235}\hat{\phi}Y_{149},
\label{okp2_24}     
\eeqa               
where $\hat{\sigma}_{\rm f235}$ is the $^{235}$U
fission cross section, $\hat{\sigma}_{\rm a}$ is the effective
neutron-absorption cross section of $^{235}$U which will be defined below,
$Y_{147}$ and $Y_{149}$ are the fission yields
of  $^{147}{\rm Sm}$ and $^{149}{\rm Sm}$, respectively, and $\hat{\phi}$
the effective flux defined in Eq.~\reflef{okp2_11}).
The effective           
neutron-absorption cross section $\hat{\sigma}_{\rm a}$
is not the bare fission cross section but an effective cross section
which includes the restitution of $^{235}$U. Restitution
refers to the sequence of events in which a neutron is captured by $^{238}$U
and $^{235}$U is the final   decay product created via successive          
 $\beta$-decays yielding $^{239}$Pu and a subsequent $\alpha$ decay.
The way we determined the effective flux
$\hat{\phi}$ and  the effective neutron-absorption cross section $\hat{\sigma}_{\rm a}$
 is given in        
Appendix B where these two quantities and the epi-thermal index are obtained
simultaneously by solving  three coupled equations. The effective neutron-absorption
cross section is defined in terms of the $^{235}$U total neutron-absorption
cross section given by
$\hat{\sigma}_{\rm 235}$  as
\beq                
\hat{\sigma}_{\rm a} = (1-C_{\rm rz})\hat{\sigma}_{\rm 235} \nonumber \\.
\eeq                
The interpretation of this definition is clear when we refer the third of the four
equations just above equation~\reflef{conc143}).
The time $t$ starts at the beginning of the reactor period, with
$\hat{\phi}$ assumed to be constant in time.\footnote{Throughout this
work, we used this assumption. However, if we use the technique used in
Ref.~[13], it is possible to remove
this assumption. Then $ \hat{\phi} t$ should be replaced $\int \hat{\phi} dt $ in all
our expressions. The numerical results given later are unchanged even if
we use this generalized expression.}
Obviously $\hat{\sigma}_{149}$ is
dominated by the thermally averaged cross section of the absorption process
\reflef{okp1_1}) while $\hat{\sigma}_{147}$ is the non-resonant absorption
cross section of $^{147}{\rm Sm}$.

The initial number densities $N_{A}(0)$ are $N^0_{235}$ for $^{235}{\rm U}$ and
\beq                
N_A(0) =N^{\rm nat}R_A^{\rm nat},
\label{okp2_23}     
\eeq                
for the samarium isotopes,  for which
the relative fractional {\em natural} abundances  $R_A^{\rm nat}$ have been
observed to be~[16]:
\beq                
R_{144}^{\rm nat} = 0.0310, \hspace{1em}R_{147}^{\rm nat} =0.1500,
\hspace{1em}R_{148}^{\rm nat} = 0.1130,\hspace{1em}\mbox{and}\hspace{1em}
R_{149}^{\rm nat} = 0.1380.
\label{okp2_23a}    
\eeq                
The value  for the stable and
inactive  $^{144}{\rm Sm}$ is included, because it is needed later, whereas
the final results will be independent of the overall normalization
constant $N^{\rm nat}$ and the capture cross section $\hat{\sigma}_{147}$.

The solution of the system of differential equations \reflef{okp2_24}) is:
\beqa               
&&N_{149}(t)=\frac{N^0_{235}\hat{\sigma}_{\rm f235}
Y_{149}}{\hat{\sigma}_{\rm a} -\hat{\sigma}_{149}} \left[ \exp
(-\hat{\sigma}_{149}\hat{\phi}t) -\exp (-\hat{\sigma}_{\rm a}\hat{\phi}t)
\right]\hspace{-.2em} +\hspace{-.2em}N^{\rm nat}R^{\rm nat}_{149}
\exp (-\hat{\sigma}_{\rm a}\hat{\phi}t),\nnb \\
&&N_{147}(t)+N_{148}(t) =\frac{N^0_{235}Y_{147}\hat{\sigma}_{\rm f235}}
{\hat{\sigma}_{\rm a}}\left[ 1- \exp (-\hat{\sigma}_{\rm
a}\hat{\phi}t)      
\right] +N^{\rm nat}\left( R^{\rm nat}_{147} + R^{\rm nat}_{148} \right),
\label{okp2_25}     
\eeqa               
from which it follows that
\beqa               
\lefteqn{\frac{N_{147}(t_1) + N_{148}(t_1)}{N_{149}(t_1)}=}\nnb\\
& & \frac{          
\frac{\begin{displaystyle}\hat{\sigma}_{\rm f235}kY_{147}\end{displaystyle}}
{\begin{displaystyle}\hat{\sigma}_{\rm a}\end{displaystyle}}
\left[ 1- \exp (-\hat{\sigma}_{\rm
a}\hat{\phi}t_1)  \right] +  \left( R^{\rm nat}_{147} + R^{\rm
nat}_{148}          
\right)}{\mbox{\Large$\frac{k\hat{\sigma}_{\mbox{\scriptsize ${\rm
f235}$}}Y_{\mbox{\scriptsize 149}}}{\hat{\sigma}_{\mbox{\scriptsize ${\rm
a}$}}-\hat{\sigma}_{\mbox{\scriptsize 149}}}$}\left[ \exp
(-\hat{\sigma}_{149}\hat{\phi}t_1) -\exp (-\hat{\sigma}_{\rm
a}\hat{\phi}t_1)  \right] + R^{\rm nat}_{149}\exp
(-\hat{\sigma}_{\rm a}\hat{\phi}t_1)},
\label{okp2_26}     
\eeqa               
at $t_1$, the end of the reactor activity. Since then the samarium isotope
concentrations $N_A(t)$ have
remained unchanged until today, unless corresponding,
naturally-occurring isotopes have flowed into the sample regions from the outside.
 
The ratio $k \equiv N^0_{235}/N^{\rm nat}$ can be determined in the following way.
We observe          
that $^{144}$Sm is never produced as a fission product and has a negligible
capture cross section, 1.64 barn~[16],  for neutron capture,
that is $N_{144}(t)$ is  constant.
We may then calculate the ratio
$N_{144}(t_1)/(N_{147}(t_1)+N_{148}(t_1))$ and compare to the observed ratio:
\beqa               
 \lefteqn{  { \left[
    \frac{N_{144}(t_1)}
         {N_{147}(t_1)+N_{148}(t_1)}
   \right]          
    }_{\rm Observed}\equiv R_{\rm O}} \nonumber \\[3ex]
& &{ \left[         
    \frac{N_{144}(t_1)}
         {N_{147}(t_1)+N_{148}(t_1)}
   \right]          
    }_{\rm Calculated} =\nonumber \\[3ex]
 & & \frac{ N^{\rm nat} R^{\rm nat}_{144} }
 { N^{0}_{235}Y_{147}
\frac{ \begin{displaystyle}\hat{\sigma}_{\rm f235}\end{displaystyle}}
{\begin{displaystyle}\hat{\sigma}_{\rm a}\end{displaystyle}}\left
[\,\,               
1 - \exp(-\hat{\sigma}_{\rm a}\phi t_1) \,\, \right]
+                   
N^{\rm nat} ( R^{\rm nat}_{147} + R^{\rm nat}_{148} ) } = \nonumber \\[3ex]
 & & \frac{R^{\rm nat}_{144} }
 { kY_{147}         
\frac{\begin{displaystyle}\hat{\sigma}_{\rm f235}\end{displaystyle}}
{\begin{displaystyle}\hat{\sigma}_{\rm a}\end{displaystyle}}
\left [\,\, 1 -     
\exp(-\hat{\sigma}_{\rm a}\phi t_1) \,\, \right]+
( R^{\rm nat}_{147} 
 + R^{\rm nat}_{148} ) }.
\label{okp2_27}     
\eeqa               
From these equations, we obtain
\beq                
k = \frac{R^{\rm nat}_{144} - R_{\rm O} ( R^{\rm nat}_{147}
 + R^{\rm nat}_{148})
}                   
{R_{\rm O}Y_{147}   
\frac{\begin{displaystyle}\hat{\sigma}_{\rm f235}\end{displaystyle}}
{\begin{displaystyle}\hat{\sigma}_{\rm a}\end{displaystyle}}\left[
\,\,1 -             
\exp(-\hat{\sigma}_{\rm a}\phi t_1) \,\,\right] }.
\label{okp2_29}     
\eeq                
which can be estimated from the observed values only.

We finally solve \reflef{okp2_26}) for $\hat{\sigma}_{149}$ by using  observed
abundances,         
and the fluence $\hat{\phi} t_1$ and the effective neutron-absorption cross
section $\hat{\sigma}_{\rm a}$
defined             
earlier             
and discussed in Appendix B.
The values of fluences and effective cross sections are tabulated in
 Ref. [17] and using them, we obtain
the results for five recently analyzed samples shown in Table \ref{table2}
\begin{table}[t]    
\bcent              
\begin{tabular}{|| c l c |c r r r r r c||} \hline
& & & & & & & & &\\[-0.07in]
& & & \multicolumn{7}{|c||}{Sample} \\[0.05in]
&  & & & SF84-1469 &SF84-1480 & SF84-1485 & SF84-1492 & SD.37& \\
& & & & & & &&& \\[-0.07in]
\hline              
& & & & & & & &&\\[-0.07in]
&$\hat{\phi}t_1$ (1/kb)&  & & 0.525 & 0.798 & 0.622 & 0.564 & 0.780&  \\
&$N_{144}(t_1)$ (\%)  &  & &  0.1052  & 0.2401 & 0.2073 & 0.1619 & 0.06909 &\\
&$N_{147}(t_1)$ (\%)  &  & & 55.34 & 53.23 & 54.03 & 54.81 & 52.74 &\\
&$N_{148}(t_1)$ (\%)  &  & & 2.796 & 3.468 & 3.079 & 2.890 & 4.694& \\
&$N_{149}(t_1)$ (\%)  &  & & 0.5544 & 0.2821 & 0.4466 & 0.4296 & 0.3088 &\\
&$N_{235}(t_1)/N_{238}(t_1)$ &  &&0.03181 & 0.02665 & 0.02971 & 0.03047 & 0.02435& \\[0.08in]
\hline              
& & & & & & & &&\\[-0.07in]
&$\hat{\sigma}_{\rm 149}$ (kb) & & &85.6 & 96.5 & 83.8 & 99.0 & 89.5 &\\[0.08in]
\hline              
\end{tabular}       
\ecent              
\caption[ctab2]{\baselineskip=12pt\small
Measured isotopic ratios and fluences for Sm for
five samples  and  corresponding,
calculated values of the cross section $\hat{\sigma}_{149}$. }
\label{table2}      
\end{table}         
.


\subsection{Isotopic composition of \label{compgd} Gd}

For gadolinium  we may define relative fractional natural abundances  $R_A^{\rm nat}$
in complete analogy with samarium in Eq.~\reflef{okp2_23}). For Gd the
relative fractional natural abundances  for
the relevant isotopes are
\beqa               
                                   R_{155}^{\rm nat} =  0.1480,
&&\hspace{1em}                       R_{156}^{\rm nat} =  0.2047,
\hspace{1em}                       R_{157}^{\rm nat} =  0.1565, \nonumber \\
& & \phantom{ xx} \\[-2ex]
\hspace{1em}                       R_{158}^{\rm nat} =  0.2484,
&&\hspace{1em}\mbox{and}\hspace{1em} R_{160}^{\rm nat} =  0.2186. \nonumber
\label{okp2_g23a}   
\eeqa               
Because             
thermal neutron capture cross section of $^{160}{\rm Gd}$ is 0.796 barn~[16],
we can use the value of the {\em nearly constant} $N_{160}(t)$
observed at each sample location to determine $N^{\rm nat}$.

Because the neutron-absorption cross sections $\hat{\sigma}_{156}$
and $\hat{\sigma}_{158}$ are
2.188 barn and 2.496 barn~[16], respectively,
 and because $^{154}$Gd is
shielded by the stable $^{154}$Sm the study of the possible shifts of the low-lying
resonances in  $^{155}{\rm Gd}$ and $^{157}{\rm Gd}$ is considerably simplified:
we may study the transmutations related to the isotope pairs
$^{155}{\rm Gd}$--$^{156}{\rm Gd}$ and
$^{157}{\rm Gd}$--$^{158}{\rm Gd}$ separately.
The differential equations  for the number of atoms
$N_{A}(t)$ per unit volume for the first isotope pair are:
\beqa               
\frac{dN_{155}(t)}{dt}&=&-\hat{\sigma}_{155}\hat{\phi}N_{155}(t) +
N^0_{235}\exp (-\hat{\sigma}_{\rm a}\hat{\phi}t) \hat{\sigma}_{\rm f235}\hat{\phi}Y_{155}, \nnb\\
\frac{dN_{156}(t)}{dt}&=&+\hat{\sigma}_{155}\hat{\phi}N_{155}(t) +
N^0_{235}\exp (-\hat{\sigma}_{\rm a}\hat{\phi}t) \hat{\sigma}_{\rm f235}\hat{\phi}Y_{156},
\label{okp2_g24}    
\eeqa               
where we ignored the small cross section $\hat{\sigma}_{156}= 2.188$~barn. We find the solution:
\beqa               
N_{155}(t)&=&\frac{N^0_{235}\hat{\sigma}_{f235} Y_{155}}{\hat{\sigma}_{\rm
a} -\hat{\sigma}_{155}} \left[ \exp (-\hat{\sigma}_{155}\hat{\phi}t)
-\exp (-\hat{\sigma}_{\rm a}\hat{\phi}t)  \right] \nnb\\
&&+N^{\rm nat}R^{\rm nat}_{155}\exp (-\hat{\sigma}_{155}\hat{\phi}t),\nnb \\
N_{155}(t)+N_{156}(t) &=&N^0_{235}\left( Y_{155} + Y_{156}
\right)\frac{\hat{\sigma}_{\rm f235}}{\hat{\sigma}_{\rm a}}\left[ 1-
\exp (-\hat{\sigma}_{\rm a}\hat{\phi}t)  \right] \nnb \\
&&+N^{\rm nat}\left( R^{\rm nat}_{155} + R^{\rm nat}_{156} \right),
\label{okp2_g25}    
\eeqa

In analogy with Eq.~\reflef{okp2_27}) we introduce
\beq                
R_{\rm O} =\frac{N_{160}(t_1)}{N_{155}(t_1) + N_{156}(t_1)},
\label{okp2_g27}    
\eeq                
which can be compared with the observed data expressed
in terms of $N_{160}/N_{156}$ and $N_{155}/N_{156}$ with reference
to the stable isotope $^{160}{\rm Gd}$.
In analogy with Eq.~\reflef{okp2_29})
the ratio $k \equiv N^0_{235}/N^{\rm
nat}$               
is here given by    
\beq                
k=\frac{R_{160}^{\rm nat} -R_0 (R^{\rm nat}_{155} + R^{\rm nat}_{156}
)}{ R_{\rm O}       
\frac{\begin{displaystyle}\hat{\sigma}_{\rm f235}\end{displaystyle}}
     {\begin{displaystyle}\hat{\sigma}_{\rm a}\end{displaystyle}}
\left( Y_{155} + Y_{156}\right) \left[  1-\exp
( -\hat{\sigma}_{\rm a} \hat{\phi}t_1)\right]   }.
\label{okp2_g29}    
\eeq                
Now $\hat{\sigma}_{155}$ can be determined by solving iteratively
\beq                
\frac{N_{155}(t_1)}{N_{156}(t_1)} =\frac{A}{B},
\label{okp2_g30}    
\eeq                
where the observed ratio is substituted into the left-hand side, and where
\beqa               
A&=& \frac{k\hat{\sigma}_{\rm f235} Y_{155}}{\hat{\sigma}_{\rm a}
-\hat{\sigma}_{155}}\left[ \exp (-\hat{\sigma}_{155}\hat{\phi}t_1 )-
\exp (-\hat{\sigma}_{\rm a} \hat{\phi}t_1 )\right] +
 R^{\rm nat}_{155}\exp (-\hat{\sigma}_{155}\hat{\phi}t_1) ,\nnb\\
B&=& \frac{k Y_{155}}{\hat{\sigma}_{155}-\hat{\sigma}_{\rm a}}
\frac{\hat{\sigma}_{\rm f235}}{\hat{\sigma}_{\rm a}}\left
(    \hat{\sigma}_{155}\left[ 1-\exp (-\hat{\sigma}_{\rm a}\hat{\phi}t_1 ) \right]
- \hat{\sigma}_{\rm a}\left[ 1-\exp (-\hat{\sigma}_{155}\hat{\phi}t_1 ) \right]\right) \nnb\\
&& \hspace{-1.5em}+kY_{156}\frac{\hat{\sigma}_{\rm f235}}{\hat{\sigma}_{\rm a}}
\left[ 1-\exp (-\hat{\sigma}_{\rm a}\hat{\phi}t_1 )
 \right] +R^{\rm nat}_{155}\left[ 1-\exp (-\hat{\sigma}_{155}\hat{\phi}t_1 ) \right] +R^{\rm nat}_{156}.
\label{okp2_g30a}   
\eeqa

The transmutation equations related to the isotope pairs
  $^{157}{\rm Gd}$ and $^{158}{\rm Gd}$ are
obtained by replacing quantities related to$^{155}{\rm Gd}$ and $^{156}{\rm Gd}$
by quantities related to
 $^{157}{\rm Gd}$ and $^{158}{\rm Gd}$, respectively, in the above equations.


\section{Analysis of the data}

 We use the isotopic data of five samples recently taken from reactor
 zones (hereafter called RZs) 10 and 13 which are located deep
 underground, 150--250 m below the surface, while the other RZs 1--9 are
 located near the surface.  The use of samples from deep-lying RZs is
 important for this study, because it minimizes contamination from
 in-flow of isotopes from the outside after the cessation of reactor
 operation.  Particularly relevant factors that influence
the contamination process are:
\begin{enumerate}   
 
 \item              
Hexavalent uranium U(VI) species are water-soluble, while
 tetravalent species are insoluble.  The valence of uranium is largely
 determined by oxidation conditions in nature.

 \item              
A uraninite sample located near the surface easily forms water-soluble species
because of contact with the oxidizing atmosphere.  On the other hand, a sample
 deep underground remains insoluble  because of a relatively reducing
environment.

 \item              
Uranium minerals under oxidizing conditions are dissolved and
recrystalized.  It is possible that rare-earth elements  originally incorporated into
uranium minerals are released  as the minerals are dissolved
and precipitated on rock surfaces.
\end{enumerate}

 It is obvious that any geological alteration of the RZs would reduce
 the quality of the isotopic data.  Rare-earth-element enriched regions were actually
 found in layers around RZs 1 to 9 [18,19].
For example, the Sm content around
 GL13 at RZ 9 shows a variation in the range from 21 to 285 ppm, a
 marked difference from the range 0.43 to 3.32 ppm observed in SF84
 which is close to RZ 10 used in this study.  This geological
 observation suggests that we as a precautionary measure, to keep the
 contamination to a minimum, collect samples from deep-lying RZs.
The quality of our new data obtained in this way  is, for example, demonstrated by a
much smaller scatter  of the values we obtained for $\hat{\sigma}_{149}$
compared to what was obtained in Ref.~[13]
from earlier samples taken from RZs 1--9.


\subsection{Description of the samples}
 
 Because near the surface, rare-earth elements and U often fail to form closed systems,
they are likely get  mixed with outside, normal isotope distributions.  This may
yield an estimate of the fluence that is either too large or
too small. We       
selected our samples in order  to minimize the effect
due to this process.   Four samples from SF84 were taken from a
 single drill core which crosscuts RZ 10 located 150 m deeper than the RZs
 near the surface.  The SD.37 sample was from a part which was highly
depleted  in        
 $^{235}$U, in a borehole which crosscuts RZ 13.  Detailed characterization of
 the reactor was performed through chemical and isotopic analyses of fission
 products and U taken every few cm from the drill core [17].

\subsection{Sm data}

\begin{figure}[t]   
\centerline{\psfig{figure=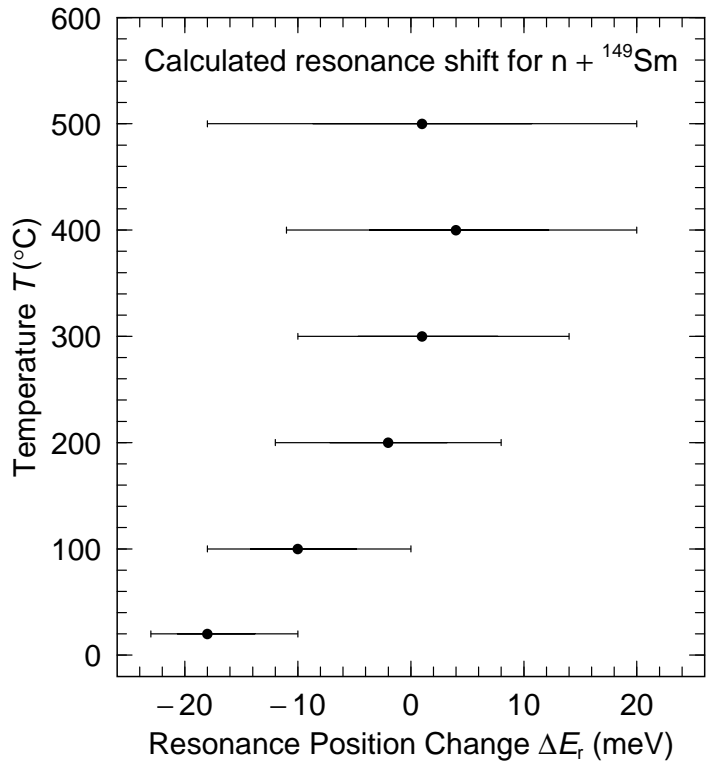,height=4.0in}}
\caption[cfig4]{\baselineskip=12pt\small
The ranges of $\Delta E_{\rm r}$, represented by horizontal lines, calculated
from Fig.~\ref{fig1} corresponding to the estimated cross section $\hat{\sigma}_{149}$ given
by (25)  are shown for several choices of the temperature, namely $T =20,$ 100,
200, 300, 400, 500$^{\circ}\mbox{C}$.}
\label{fig4}        
\end{figure}

Table \ref{table2} shows measured and calculated results for the five
samples labeled in the first row. The first
four samples were taken from RZ 10, whereas the last sample was
taken from  RZ 13.  
Measurement errors are given in Ref.~[17]. Because they are mostly
below the level of percent they are negligible in our analysis.
More details on this question will be reported elsewhere.
The last row shows $\hat{\sigma}_{149}$ calculated as
described in Subsection 3.1.  The result can be represented by
\beq                
\hat{\sigma}_{149} =91 \pm 6\mbox{ kb}.
\label{okp3_1}      
\eeq                
We then use Fig.~\ref{fig1} to estimate $\Delta E_{\rm r} = E_{\rm r}-  E_{\rm
r0}$, for the assumed temperatures, with the results shown in Fig.~\ref{fig4}.

Note that for the temperature range of interest there are two
separated regions for the allowed values  $\Delta E_{\rm r}$ for a given set of
$\hat{\sigma}_{149}$ and the lower and the upper temperature
limits $T_{\rm lower}$ and $T_{\rm upper}$.  In Fig. 1, these regions
are bounded by both ends of \reflef{okp3_1}), $T_{\rm lower}$ and  $T_{\rm upper}$, and the
left-branch of the curves for one while the right-branch for the other.  The
right-branch range passes through the area that covers $\Delta E_{\rm
r} =0$ whereas  the left-branch range extends to the far-negative
region of $\Delta E_{\rm r}$.

By taking advantage of the strong temperature-dependence 
of the neutron-capture cross section of $^{176}{\rm Lu}$, 
most estimates of the reactor temperature have been made based 
on observed isotopic ratios of $^{176}{\rm Lu}/^{175}{\rm Lu}$.  
It has been suspected, however, that the result is highly problematic.
In one of the careful analyses the temperature range $200-360
^{\circ}{\rm C}$ was obtained [20], whereas the same technique
 applied to another set of samples produced  $380^{\circ}{\rm C}$ for
one sample  with all the six others yielding $> 1000^{\circ}{\rm
C}$ [17], indicating still unknown flaws, such as the lack of more precisely
determined temperature-dependence of the capture cross section
particularly at higher temperatures, possible outside contamination, and
others, depending on different conditions of reactors.

In view of this complicated and even confusing situation, it might be
worth looking for other
types of approach.  A range of 300--350$^{\circ}{\rm C}$ was obtained,
for example, by measuring the oxygen isotopes of chlorites [21]
outside reactor cores, though less direct.  This is supported by the
 most recent analysis by Pourcelot and Gauthier-Lafaye who have
obtained   the reactor temperature from  the crystallization
temperature of chlorites in the reactors [22].  Based on this result,
and by considering other possible factors, like distance to the
reactor core and the cooling efficiency,  the authors concluded
200--400$^{\circ}$C, which we adopt here as the most reliable
temperature range currently available.  Notice that this range is
consistent with the Lu result in Ref. [20].

We then obtain 
\beq
\Delta E_{\rm r} = 4\pm 16\;{\rm meV},\quad\mbox{for}
\quad T=200\mbox{--}400^{\circ}{\rm C},
\label{okp3_2}
\eeq
 for the right-branch range, while
 \beq                
 \Delta E_{\rm r} = 
  -97\pm 8\;{\rm meV},\quad\mbox{for}\quad T=200\mbox{--}400\;^{\circ}{\rm C},
 \label{okp3_2c}
 \eeq 
for the left-branch range.  The standard deviation has been determined from 
the lower and upper limits  of the temperature.

We also notice that restricting the temperature range further to
$300\mbox{--}400\;^{\circ}{\rm C}$, for examples, as suggested by
Refs. [20,22], does not improve the result
\reflef{okp3_2}) in any
significant manner;
\beq
\Delta E_{\rm r} = 5\pm 16\;{\rm meV},\quad\mbox{for}\quad T=300\mbox{--}400^{\circ}{\rm C}.
\label{okp3_2a}
\eeq

We point out that the shift as large as $\sim -100{\rm meV}$
suggested in
\reflef{okp3_2c}) would imply
 a level even below the threshold, hence requiring a more careful
 analysis.  It seems still unlikely that it would change
 substantially the claim that $E_{\rm r}$ at the time of the natural
 reactor was different from the present value beyond many standard
 deviations.

The bound thus obtained for $\Delta E_{\rm r}$, particularly for the right-branch, are rather insensitive to an increase of $T_{\rm upper}$ as
long as $T_{\rm upper} \lsim T_{\rm crit} \approx 820^{\circ}{\rm C}$, a critical temperature above which the two regions are bridged to each other.
In fact for $T_{\rm upper} = 700^{\circ}{\rm C}$, for example, we find $-83\pm
22{\rm meV}$, and $11\pm 10{\rm meV}$ for the left-branch and the
right-branch ranges, respectively.  The separation between the centers
of the two ranges is 94meV, which is nearly three times as large compared with
the sum 32meV of the half-widths of each region.  Even if we take the
two standard deviation bound for $\hat{\sigma}_{149}$, the critical
temperature is estimated to be $T_{\rm crit} \approx 730^{\circ}{\rm
C}$, and the ranges for $T_{\rm upper} = 700^{\circ}{\rm C}$ are found
to be $-78\pm 30{\rm meV}$ and $6\pm 26{\rm meV}$, respectively, which
are still well separated, with the  separation 84meV and the sum of
the half-widths 56meV.

It seems also useful to note that assuming temperatures $\gsim
1000^{\circ}{\rm C}$ leaves little room for $\Delta E_{\rm r}$ in
Fig.~\ref{fig1} for \reflef{okp3_1}), making higher temperatures less
likely from the point of view of theoretical analyses.

As we alluded to before, one may suspect that  the samples
have been subjected to some  post-reactor
contamination.  To estimate  this  effect,
we assume that a fraction $\zeta$ of the natural abundance
$N^{\rm nat}$(see \reflef{okp2_23a})) flowed into
the reactor from the outside
environment, contributing to the observed amount of Sm.
We subtract this inflow from the now observed isotope abundances and
apply the preceding analysis to the {\em remainder} to re-calculate
$\hat{\sigma}_{149}$.  As we can see in Table
\ref{table2a}, the result is quite insensitive to $\zeta$ as
\begin{table}[tb]   
\bcent              
\begin{tabular}{||clc |c r r r r r c||} \hline
&&&&&&&&& \\[-0.07in]
& $\zeta$ &&& SF84-1469 &SF84-1480 & SF84-1485 & SF84-1492 &
\hspace{1.4em} SD.37 &\\
&&&&&&&&& \\[-0.07in]
\hline              
&&&&&&&&& \\[-0.07in]
& 0.00 &&& 85.6  & 96.5 & 83.8& 99.0 & 89.5 &\\
& 0.01  &&   & 86.4 & 100.2 & 85.6 & 100.7 &  90.4 &\\
&0.02  & & & 87.1  & 104.3 & 87.4 & 102.4 & 91.3  &\\
&0.03  &  && 87.8 & 108.6 & 89.3 & 104.2 & 92.2 &\\
&0.04  &&  & 88.6 & 113.4 & 91.2 & 106.0 & 93.2 &\\[0.08in]
\hline              
\end{tabular}       
\ecent              
\caption[ctab2a]{\baselineskip=12pt\small
Estimated cross sections $\hat{\sigma}_{149}$ (kb) for five values of the
post-reactor contamination $\zeta$  for five samples. }
\label{table2a}     
\end{table}         
long as it stays around a few percent. This contamination range is
suggested  by our analysis of Gd below.

For $\zeta = 0.04$, for example, we obtain
\beq                
\hat{\sigma}_{149} =99 \pm 10\;  {\rm kb}
\label{okp3_1a}     
\eeq                
instead of \reflef{okp3_1}), from which follows
\beq                
\Delta E_{\rm r} = 2\pm 14\;  {\rm meV},
\label{okp3_2b}     
\eeq                
for either of the temperature ranges
200--400$^{\circ}{\rm C}$ and
300--400$^{\circ}{\rm C}$.  Although the
 result \reflef{okp3_1a}) seems worse than \reflef{okp3_1}),
the result \reflef{okp3_2b}) for $\Delta E_{\rm r}$ puts an even stronger limit
on its time-variation
than \reflef{okp3_2}), because the curves corresponding to different temperatures
in Fig.~\ref{fig1} happen to all intersect  near $\Delta E_{\rm r}=0$.

 
\subsection{Gd data}
 
In analogy with our studies of the Sm
isotopic compositions and estimates
of neutron-capture cross sections tabulated in Table
\ref{table2} we now study the
measured Gd isotopic compositions in the Oklo RZs, which are  tabulated
in Table \ref{table3}.  To obtain this table, we used the fluence and effective
cross-section values which are tabulated in
[17] in the same way as we did for Sm case. By use of the formalism developed in Section
\ref{compgd} we obtain the estimates for the neutron-capture
cross sections listed in the bottom two lines of Table \ref{table3}.
We find that   the resonance-position change $\Delta E_{\rm r}$ deduced from
the cross sections  in  Table \ref{table3} are
\begin{table}[tb]   
\bcent              
\begin{tabular}{||clc |c r r r r r c||} \hline
&&&&&&&&& \\[-0.07in]
& & & \multicolumn{7}{|c||}{Sample} \\[0.05in]
&  &&& SF84-1469 &SF84-1480 & SF84-1485 & SF84-1492 & SD.37 &\\
&&&&&&&&& \\[-0.07in]
\hline              
&&&&&&&&& \\[-0.07in]
& $N_{155}(t_1)$(\%)  &&   & 0.5006  & 0.4608 & 0.6065 & 0.3899 & 0.5915 &\\
& $N_{156}(t_1)$(\%)  &&   & 30.03 & 29.46 & 30.79 & 29.90 & 30.37  &\\
& $N_{157}(t_1)$(\%)  &&   & 0.0418  & 0.2641 & 0.1921 & 0.0505 & 0.1881 &\\
& $N_{158}(t_1)$(\%)  &&   & 23.66 & 25.07 & 26.52 & 23.68 & 17.31 &\\
& $N_{160}(t_1)$(\%)  &&   & 11.21 & 12.24 & 12.93 & 10.95 & 6.1030 &\\[0.08in]
\hline              
&&&&&&&&& \\[-0.07in]
 
& $\hat{\sigma}_{\rm 155}$(kb)&& & 30.9 & 16.8 & 17.8 & 36.7 & 26.3 &\\
& $\hat{\sigma}_{\rm 157}$(kb) &&& 83.3 & 8.0 & 14.3 & 73.7 & 23.3 &\\[0.08in]
\hline              
\end{tabular}       
\ecent              
\caption[ctab3]{\baselineskip=12pt\small
Measured isotopic rations  for Gd for  five samples and the corresponding,
calculated cross sections $\hat{\sigma}_{155}$ and $\hat{\sigma}_{157}$.  }
\label{table3}      
\end{table}         
generally significantly
\begin{table}[tb]   
\bcent              
\begin{tabular}{||c l c | c r r c  r r c||} \hline
&  &  &  &                               & & & & & \\[-0.07in]
&  &  &  & \multicolumn{2}{c}{SF84-1469}
         & \mbox{}\hspace{1em}\mbox{}
         & \multicolumn{2}{c}{SF84-1492} &\\
   & \multicolumn{1}{c}{$\zeta$} &
&  & \multicolumn{1}{c}{$\hat{\sigma}_{155}$}
   & \multicolumn{1}{c}{$\hat{\sigma}_{157}$}
&  & \multicolumn{1}{c}{$\hat{\sigma}_{155}$}
   & \multicolumn{1}{c}{$\hat{\sigma}_{157}$} & \\
&  &                
&  & \multicolumn{1}{c}{(kb)}
   & \multicolumn{1}{c}{(kb)}
&  & \multicolumn{1}{c}{(kb)}
   & \multicolumn{1}{c}{(kb)}                 & \\[0.08in]
\hline              
&  &  &  &                               & & & & & \\[-0.07in]
& 0.000 & & & 30.9  &  83.3 & & 36.7 &  73.7  &\\
& 0.001 & & & 31.4  & 102.9 & & 37.4 &  87.2  &\\
& 0.002 & & & 31.9  & 134.7 & & 38.1 & 106.6  &\\
& 0.003 & & & 32.3  & 195.0 & & 38.8 & 137.1  &\\[0.08in]
\hline              
\end{tabular}       
\ecent              
\caption[ctab3a]{\baselineskip=12pt\small
Calculated cross sections $\hat{\sigma}_{155}$  and
$\hat{\sigma}_{157}$  for
two samples and four values of  the post-reactor contamination $\zeta$. }
\label{table3a}     
\end{table}         
different from zero in contrast to what
\begin{table}[tb]   
\bcent              
\begin{tabular}{|| c l c | c r r  c r r  c r r  c ||} \hline
&  &  &  &                               & & & & & & & & \\[-0.07in]
&  &  &  & \multicolumn{2}{c}{SF84-1480}
         &\mbox{}\hspace{1em}\mbox{}
         &\multicolumn{2}{c}{SF84-1485}
         &\mbox{}\hspace{1em}\mbox{}
         &\multicolumn{2}{c}{SD.37}           & \\
& \multicolumn{1}{c}{$\zeta$} &
&  & \multicolumn{1}{c}{$\hat{\sigma}_{155}$}
   & \multicolumn{1}{c}{$\hat{\sigma}_{157}$}
&  & \multicolumn{1}{c}{$\hat{\sigma}_{155}$}
   & \multicolumn{1}{c}{$\hat{\sigma}_{157}$}
&  & \multicolumn{1}{c}{$\hat{\sigma}_{155}$}
   & \multicolumn{1}{c}{$\hat{\sigma}_{157}$} & \\
&  &                
&  & \multicolumn{1}{c}{(kb)}
   & \multicolumn{1}{c}{(kb)}
&  & \multicolumn{1}{c}{(kb)}
   & \multicolumn{1}{c}{(kb)}
&  & \multicolumn{1}{c}{(kb)}
   & \multicolumn{1}{c}{(kb)}                 & \\[0.08in]
\hline              
&  &  &  &                               & & & & & & & & \\[-0.07in]
& 0.00 & & & 16.8 &   8.0 & & 17.8 &  14.3 && 26.3&  23.3  &\\
& 0.01 & & & 20.2 &  11.1 & & 20.6 &  26.4 && 28.1&  29.9  &\\
& 0.02 & & & 25.4 &  20.6 & & 24.4 & 236.9 && 30.2&  42.0  &\\
& 0.03 & & & 34.4 & 227.1 & & 30.1 &   --- && 32.7&  70.9  &\\
& 0.04 & & & 53.5 &  ---  & & 39.2 &   --- && 35.6& 235.3  &\\[0.08in]
\hline              
\end{tabular}       
\ecent              
\caption[ctab3b]{\baselineskip=12pt\small
Calculated cross sections $\hat{\sigma}_{155}$ and $\hat{\sigma}_{157}$
for three samples and five values of the post-reactor contamination $\zeta$.
Dash (---) entries are given when
the calculated cross sections are much larger than any of those occurring in
 Fig.~\ref{fig3}.}  
\label{table3b}     
\end{table}         
was indicated by the  Sm results.  It is therefore natural to investigate if
the effect of contamination here, for Gd, plays a more crucial role than
for Sm, which we found was relatively insensitive to modest contaminations.
For Gd we           
divide the five samples into two different
classes, which are presented separately in Tables \ref{table3a} and \ref{table3b}.
The samples SF84-1469 and SF84-1492 presented in Table \ref{table3a} are characterized by
 much smaller abundances of $^{157}{\rm Gd}$ than are the other three samples
presented           
in Table \ref{table3b}.

We assume that  for each sample
the {\em unknown} contamination parameter $\zeta$ is the same for
$^{155}{\rm Gd}$ and $^{157}{\rm Gd}$, and investigate if there is
any $\zeta$ for which we obtain the same $\Delta E_{\rm r}$ for both
isotopes.  This is based on the assumption that the mass scale ${\cal
M}$ that relates $\Delta E_{\rm r}$ to $\Delta\alpha$  are the same
for the two isotopes.  In fact ${\cal M}$ defined by (see
\reflef{okp4_1}) later)
\beq
\alpha\frac{d \Delta E_{\rm r} }{d\alpha} ={\cal M},
\label{okp3_5} 
\eeq
is found to be nearly common ($\sim -1.1 {\em {\rm MeV}}$) according
to  a phenomenological Bethe-Weizs\"{a}cker formula, for example.  
  Although the result might be corrected roughly by a factor
of two if a more precise measurement is used [13], we
consider the above simple estimate to be sufficient for our present
purpose.  In the same context, it seems reasonable to assume that
${\cal M}$ is nearly the same also for Sm, at least sharing the same
sign.

\begin{figure}[t]   
\centerline{\psfig{file=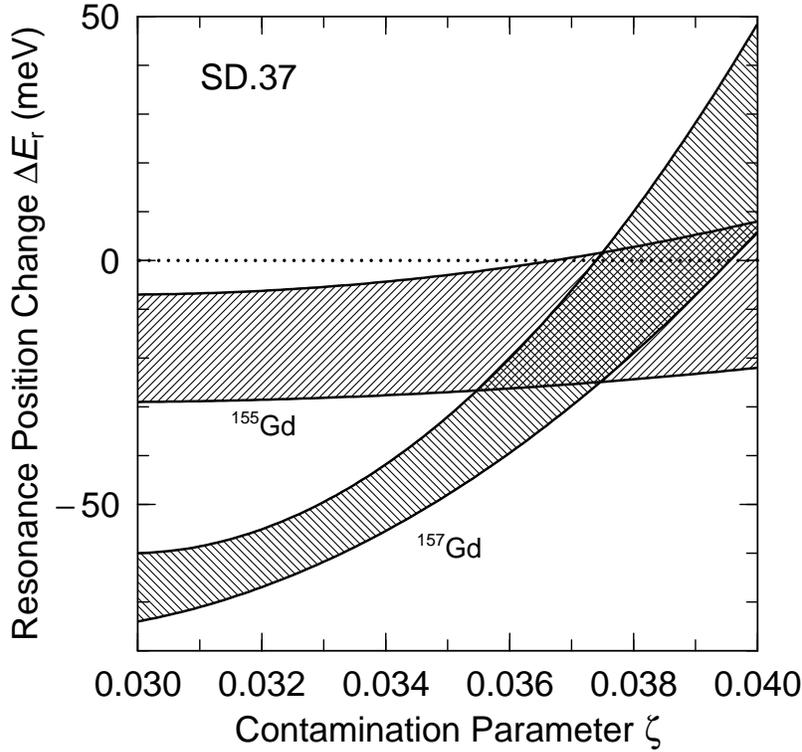,height=4.0in}}
\caption[cfig5]{\baselineskip=12pt\small
For the sample SD.37 we plot the resonance-position change $\Delta E_{\rm r}$ for
$^{155}{\rm Gd}$ and $^{157}{\rm Gd}$ versus  the contamination parameter $\zeta$, the
ratio of the post-reactor contamination to the natural abundance.
The two bands show the range of $\Delta E_{\rm r}$ for the temperature
range 200--400$^{\circ}$C, with the lower bound of each band corresponding to
$T=200$$^{\circ}$C and the upper bound to  $T=400$$^{\circ}$C.
From the location of the overlap region of the two bands  we deduce
that the contamination parameter $\zeta$ is in the range 0.0355 to 0.0403
and that the resonance-position change
$\Delta E_{\rm r}$ is in the range $-26$ meV to $+9$ meV.  }
\label{fig5}        
\end{figure}

For the sample SD.37 we have in  Fig.~\ref{fig5} plotted $\Delta
E_{\rm r}(\zeta)$ determined in the following way.
The neutron-capture cross sections $\hat{\sigma}_{155}$ and
$\hat{\sigma}_{157}$ corresponding to the observed isotopic distributions
with the assumed contamination removed, are
calculated.         
The resonance-position changes corresponding to these cross sections are then
obtained for the assumed reactor temperature limits from  the appropriate curves in
Figs.~\ref{fig2} and \ref{fig3}.

Like in Fig.~\ref{fig1}, we have also two ranges.  It is now the left-branch
ranges that passes through the area covering $\Delta E_{\rm r}=0$.
For the right-branch ranges we obtain large {\em positive} values
for $\Delta E_{\rm r}$. This branch is not  consistent
with the Sm data giving either a near-zero  branch or a negative
branch range for $\Delta
E_{\rm r}$. The requirement that the Sm and Gd data give
the same result for the time variability of $\alpha$ therefore only
allows the near-zero branches for Sm and Gd as acceptable choices 
for $\Delta E_{\rm r}$.

For the assumed temperature range
200--400$^{\circ}$C, the bands for $^{155}{\rm Gd}$
and $^{157}{\rm Gd}$ corresponding to the left-branch ranges do intersect  each other for $\zeta$ in the range 0.036--0.040, yielding the range
$-26$ meV to $+9$~meV  for $\Delta E_{\rm r}$, as shown in
Fig.~\ref{fig5}.  We  also find that the band for $^{155}{\rm Gd}$ is
rather flat whereas that for $^{157}{\rm Gd}$ shows a rapid increase
as $\zeta$ changes.  For this reason the common intersection gives
$\Delta E_{\rm r}$ staying rather close to zero, even if the assumed
equality for the two $\Delta E_{\rm r}$'s is somewhat modified.  Consider the equality is changed
by a factor two, for example, in either direction.  We find that the
common range for $\zeta$ stays near $3\sim 4$\%, while that for $\Delta
E_{\rm r}$ is  expanded nearly twice as large, still leaving the
conclusion of the consistency with the Sm data basically unchanged.

Similar results are
obtained for the  other samples except for SF84-1492 for which no
reasonable solution is obtained.

\begin{figure}[tbh] 
\centerline{\psfig{figure=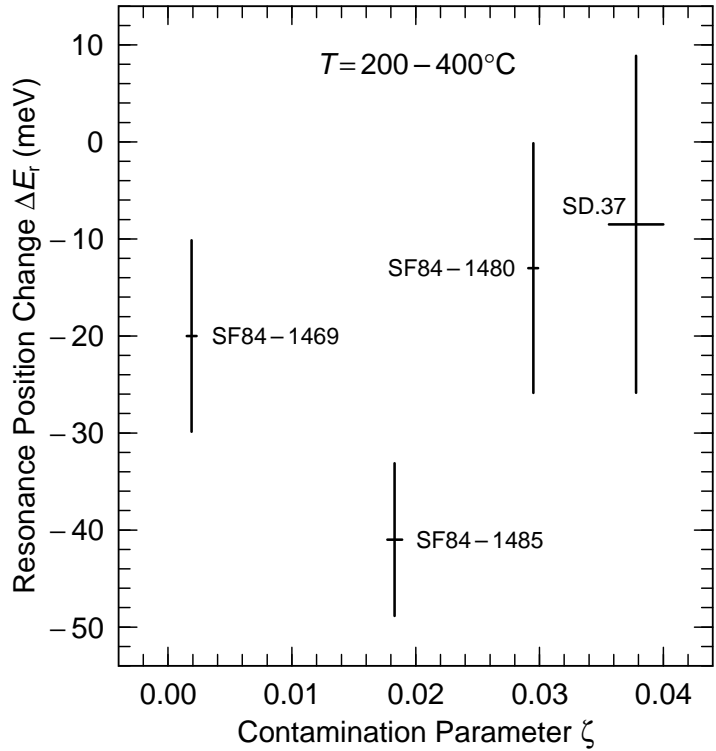,height=4.0in}}
\caption[cfig6]{\baselineskip=12pt\small
Summary of the ranges of $\Delta E_{\rm r}$
and $\zeta$ estimated from  Gd data from four samples, for
the assumed temperature range $T=200$--400$^{\circ}$C.
}                   
\label{fig6}        
\end{figure}

In Fig.~\ref{fig6} we summarize the
ranges of $\zeta$ and $\Delta E_{\rm r}$ obtained for the four samples.
It is               
remarkable to find that, with the  exception of SF84-1485, the
obtained values of $\Delta E_{\rm r}$ are consistent with the Sm
result \reflef{okp3_2}).  The presence of contamination is clearly
indicated and its magnitude is also consistent with the Sm data.

As for SF84-1485 and SF84-1492, one may suspect non-uniform contamination  inside
the reactor core or some other type of error.  This suggests that
more stringent limits on the variation of $\Delta E_{\rm r}$
would be achieved by collecting additional samples.  It might
be argued that assuming a common value of the contamination parameter $\zeta$ for the
two isotopes is not well founded.  However, it is the most simple
assumption,         
and with this assumption the Sm and Gd results are entirely consistent
with each other,    
both for the contamination parameter $\zeta$ and for
the resonance-position change $\Delta E_{\rm r}$ for which the Sm and
Gd mutually support           
the conclusion  $\Delta E_{\rm r}=0$, though the conclusion is still
tentative before the above plausibility assumptions are justified by
more careful considerations, taking the need for more samples aside.

We also notice that, unlike in the Sm data, the result on Gd depends on the temperature rather sensitively.  If we assume twice as wide the temperature range $200-600^{\circ}$C, for example, the latitude of $\Delta E_{\rm r}$ in Fig. 6 extends roughly in the same rate.  This tends to weaken the consistency check between the null results of Sm and Gd, but not to the extent that the non-null results are allowed to be mutually consistent.  From this point of view, further improving the temperature estimate is strongly encouraged.

As an additional remark about the temperature, the same argument as in Sm showing that temperatures higher than $1000^{\circ}$C tend to constrain the cross sections to too small values compared with the measured values applies to Gd as well, though the result ought to be more complicated.


\section{Relation of the resonance-position change to the variation of the
coupling constants}

Following Damour and Dyson we consider $\Delta E_{\rm r}$ as the
change of the mass scale ${\cal M}_{\rm c}$ representing the difference of the expectation values 
of the Coulomb energy $H_{\rm c}$;
\beq
{\cal M}_{\rm c} = <H_{\rm c}>_1 -<H_{\rm c}>_2,
\label{okp4_1a} 
\eeq
where the subscripts 1 and 2 refer to the excited state of
$^{150}{\rm Sm}$ and the ground state of $^{149}{\rm Sm}$ plus a 
neutron, respectively.  They estimated ${\cal M}_c\approx -1.1 {\rm MeV}$, which is
proportional to $\alpha$ so that [13]  
\beq                
\frac{\Delta {\cal M}_{\rm c}}{{\cal M}_{\rm c}}= \frac{\Delta\alpha}{\alpha}.
\label{okp4_1}      
\eeq

Identifying $\Delta {\cal M}_{\rm c}$ with \reflef{okp3_2}) we obtain
\beq                
\frac{\dot{\alpha}}{\alpha} =(-0.2 \pm 0.8)\times 10^{-17}\;{\rm y}^{-1},
\label{okp4_6}      
\eeq 
which is about five times more stringent than in Ref.~[13].  The improvement is
 mainly due to our choice of the separated right-branch range rather
than the combined range.

If, on the other hand,  we choose
\reflef{okp3_2c}) corresponding to the left-branch range, we would obtain 
 \beq
 \frac{\dot{\alpha}}{\alpha}= (4.9 \pm 0.4)
  \times 10^{-17}\;{\rm y}^{-1},
 \label{okp4_6_2}
 \eeq
indicating an apparent evidence of the time-variability of $\alpha$.

We note that  $E_{\rm r}=97.3 \rm {meV}$ of the resonance in
question is extremely small compared with $|{\cal M}_{\rm c}| \approx
1 {\rm MeV}$, most part of which is cancelled by the contribution from
the strong interaction ${\cal M}_{\rm s}$ defined analogously with
${\cal M}_{\rm c}$;
\beq
{\cal M}_{\rm c} \approx -{\cal M}_{\rm s}.
\label{okp4_6_2a}
\eeq
We assume ${\cal M}_{\rm s}$ is proportional to $\alpha_{\rm s} ^n$ with $n$
likely to be 1, but allowing $n\neq 1$ anticipating some complication of
nuclear physics.  Combining this with \reflef{okp4_1}) yields
\beq
\frac{\Delta\alpha_{\rm s}}{\alpha_{\rm s}}\approx -\frac{1}{n}\frac{\Delta\alpha}{\alpha},
\label{okp4_6_2b}
\eeq 
if the change of the resonance energy is solely due to the change of the 
strong interaction coupling constant.  In this way we expect simply
the minus of \reflef{okp4_6}) and \reflef{okp4_6_2}) for the assumed
value $n=1$.  These are somewhat weaker than what we might find by
assuming ${\cal M}_{\rm s} \approx  40 {\rm MeV}$, as suggested originally
by  Shlyakhter, who naively ignored  the fact that the ground state
$^{149}{\rm Sm}$ has its own binding energy, which should be crucial for the near cancellation between the Coulomb and nuclear energies.

In principle, one might argue that the nuclear energy levels should be
expressed in terms of the more fundamental constant $\alpha_{\rm QCD}$.  When applied to realistic low-energy
phenomenology, however, it does not appear that this approach 
has reached the same level of rigor as QED.  We instead expect to
obtain a useful insight by expressing the result on the energy shift
in terms of $\alpha_{\rm s}$, which has played important roles widely in understanding low-energy nuclear physics.

On the other hand, the argument can also be extended to a perhaps more realistic situation in which $\Delta E_{\rm r}$ is a combined effect both of the Coulomb and strong interactions.  To show this we rewrite \reflef{okp4_6_2a}) without omitting $E_{\rm r}$;
\beq
{\cal M}_{\rm c} +{\cal M}_{\rm s} =E_{\rm r}.
\label{okp4_6_2c}
\eeq
We also assume $n=1$, for the moment.  By varying $\alpha$ and $\alpha_{\rm s}$ in \reflef{okp4_6_2c}), we obtain
\beq
\frac{\Delta\alpha}{\alpha}{\cal M}_{\rm c} +\frac{\Delta\alpha_{\rm
s}}{\alpha_{\rm s}}{\cal M}_{\rm s} =\Delta E_{\rm r}.
\label{okp4_6_2d}
\eeq
This is not enough to determine $\Delta\alpha /\alpha$ or
$\Delta\alpha_{\rm s} /\alpha_{\rm s}$ separately.  We need another relation
between the two quantities.  Here we simply introduce the ratio $u=(\Delta\alpha_{\rm s} /\alpha_{\rm s})/ (\Delta\alpha /\alpha)$ as a convenient parameter, in terms of which we solve \reflef{okp4_6_2d}) for $\Delta\alpha /\alpha$;
\beq
\frac{\Delta\alpha}{\alpha}=\frac{\Delta E_{\rm r}}{(1-u){\cal M}_{\rm 
c} +uE_{\rm r}}.
\label{okp4_6_2e}
\eeq
Unless $u$ is very close to 1, we find
\beq
(1-u)\frac{\Delta\alpha}{\alpha}=\left( \frac{1-u}{u}\right)
\frac{\Delta\alpha_{\rm s}}{\alpha_{\rm s}}= \frac{\Delta E_{\rm
r}}{{\cal M}_{\rm c}},
\label{okp4_6_2f}
\eeq
hence giving $\dot{\alpha}/\alpha$ essentially of the same size as \reflef{okp4_6}) or \reflef{okp4_6_2}).

If, however, 
\beq
u-1 \ll \frac{E_{\rm}}{{\cal M}_{\rm c}}\sim {\cal O}\left( 10^{-7} \right),
\label{okp4_6_2g}
\eeq
the denominator in \reflef{okp4_6_2e}) is almost $E_{\rm r}$, giving
\beq
\frac{\Delta\alpha}{\alpha}= \frac{\Delta\alpha_{\rm s}}{\alpha_{\rm s}}=\frac{\Delta E_{\rm r}}{E_{\rm r}},
\label{okp4_6_2h}
\eeq
which could be as large as 0.2, resulting in $\dot{\alpha}/\alpha$ as large as $10^{-10}{\rm y}^{-1}$.  The condition \reflef{okp4_6_2g}) implies, however,  an unrealistic fine-tuning of parameters of any theoretical models.

\section{Discussions}

We have re-examined in detail the derivation of the upper 
bound on the shift with time of the resonance energy of $^{149}{\rm Sm}$,
 and applied it to the new, carefully collected samples from
deep-lying Oklo reactor zones.  We reached the conclusion $|\Delta
E_{\rm r}| \lsim 16$ meV.  The surprising agreement to Shlyakhter's
original result is probably somewhat accidental, since he in early
years probably had access to  few uncontaminated samples 
 and since he also assumed the unrealistically low temperature $T=20^{\circ}$C.  

We  still have another choice of nonzero shift $\Delta E_{\rm r}\sim
-100{\rm meV}$.  From the former more favored choice,  we  derived 
$|\dot{\alpha} /\alpha |\approx |\dot{\alpha}_{\rm s} /\alpha_{\rm s}|\lsim
 10^{-17}\;{\rm y}^{-1}$, unless there is any reason why the equality
$\dot{\alpha} /\alpha =\dot{\alpha}_{\rm s} /\alpha_{\rm s} $ holds
true to an extreme accuracy of $10^{-7}$.  The near equality for the
time-variability of the two coupling constants is a consequence of the near cancellation between the two interactions resulting in an extremely
small value of the resonance energies.

The cross section deduced from the 15 samples analyzed in
Ref.~[13]  is $\hat{\sigma}_{149}= 75\pm 18$~kb (two
standard deviation).    We find that the corresponding critical
temperature for joining two ranges is $T_{\rm crit} \approx
900^{\circ}{\rm C}$, which is sufficiently higher than their choice $T_{\rm upper} =700^{\circ}{\rm C}$, implying that the two regions can be considered well separated.  In this sense we may interpret their result in terms of the two ranges;
\beq
-94\pm 26 {\rm meV},\quad\mbox{and}\quad 46\pm 44 {\rm meV}.
\label{e1}
\eeq
The separation and the sum of the half-widths are 140meV and 70meV,
respectively. The authors seemed nevertheless  to  favor an even more conservative attitude to consider as if the two ranges were connected, the far ends of which give 
\beq
-120{\rm meV} < \Delta E < 90 {\rm meV}.
\label{e2}
\eeq
The limits might be narrowed slightly to $-115$ meV and $68$ meV, respectively, at the level of one standard deviation, to be compared with our analysis.
Notice also that $\Delta E_{\rm r}=0$ lies just on the edge of the two-standard deviation bound in the second of
(\ref{e1}), indicating probably contamination of their samples.

We also studied the isotopic composition of Gd in Oklo, to look, for the first
time, for a possible time-variation in the position of the low-lying
resonances in $^{155,157}{\rm Gd}$ and a corresponding time-variation
of the coupling constants. We used the same methodology as in the studies of Sm. In the Gd analysis it was crucial to determine the amount of  post-reactor 
contamination, in contrast to the case in the Sm analysis.  By
assuming that the energy-shifts of both isotopes of Gd are nearly the
same having the same sign as that of Sm, the results of the Gd studies confirm and reinforce the conclusion of the null result 
based on the $^{149}{\rm Sm}$ data.  To consolidate this tentative but 
reasonable conclusion, further analyses might be required.  This
should include, in addition to collecting more samples, re-examining
the proportionality between ${\cal M}_{\rm c}$ and $\alpha$ as assumed in [13], because any change in the Coulomb energy affects the size of a nucleus hence changing the kinetic energy as well, resulting in rather complicated dependence of the {\em total} energy on $\alpha$ in general.  This aspect might be even more relevant when we add the contribution from the strong interaction.

We hope that we have
shown with our detailed analysis here, having  somewhat improved the result due to Damour and Dyson, that
the constraints from the Oklo phenomenon should be considered 
to be quite  realistic upper bounds on the time-variability of the fundamental constants.
This result should provide valuable guidance in the search for viable unification models.

The result obtained here, that the upper bounds on the rate-of-change of the fundamental constants 
are much lower  
than the value $\sim 10^{-10}\;{\rm y}^{-1}$ 
expected naturally in terms of the present age of the universe is likely linked to the
cosmological constant problem, suggesting the {\em presence} of a
small but nonzero $\Lambda$ [23-27].
One of the unique features of the approach  in this paper, in contrast
to the cosmological studies of the time-variation of the fundamental
constants, is that we probe the time
$\sim 2\times 10^{9}$ y ago.  This is somewhat less distant than many of the 
cosmological phenomena considered so far, which have 
typical time scales of  $\sim 10^{10}$ y [3,5,8].
The different time scales of the studies  may be useful in trying to
understand the recent finding that the
fine-structure constant seems to exhibit a complicated time dependence
 [28].  
\bigskip

We thank John Barrow for giving us  important insight to the
whole subject through his book [29] and for providing  
access to Ref. [7].
We are grateful to Thibault Damour for pointing out to us the
importance of the effective cross section in the analysis of the  Oklo phenomena.

The present work was partially supported by the REIMEI Research Resources of 
Japan Atomic Energy Research Institute and by the US Department of Energy,
 also in part by Grant-in--Aid for Scientific Research
of the Ministry of Education, Science and Culture.

\bigskip            
\newpage            
\bcent              
{\Large\bf Appendices}
\ecent              
\renewcommand{\theequation}{\Alph{section}.\arabic{equation}}
\setcounter{equation}{0}
\appendix

\section{Numerical details}

The integral        
\beq                
I_c = \int_0^{\infty} dE \sqrt{E}\exp
(-E/T)\frac{E_{\gamma}}{(E-E_{\rm r})^2 +\Gamma_{\rm tot}^2 /4}
\label{okpa_1}      
\eeq                
occurring in Eq.~\reflef{okp2_9}) has to be evaluated
by numerical quadrature.
So that we can  use  Gauss-Laguerre integration we make the substitution
$x=E/T$ and obtain  
\beq                
I_c = T \int_0^{\infty} dx \sqrt{xT}\exp (-x) \frac{E_{\gamma}}{ (xT-E_{\rm r}
)^2 + \Gamma_{\rm tot}^2 /4} = T \int_0^{\infty} dx \exp (-x) f_1(x).
\label{okpa_2}      
\eeq                
where               
\beq                
f_1(x) =\sqrt{xT}  \frac{E_{\gamma}}{ (xT-E_{\rm r}
)^2 + \Gamma_{\rm tot}^2 /4}.
\label{okpa_3}      
\eeq                
Because $f_1$ has a maximum at
\beq                
x_{\rm max}=\frac{ E_{\rm r} + \sqrt{4E_{\rm r}^2 + 3 (\Gamma_{\rm
tot}/2)^2 } }{4 T}, 
\label{okpa_4}      
\eeq                
it is necessary to divide the
integration interval into two parts to achieve satisfactory numerical
accuracy. The first part of the integration should include the maximum
of  the function $f_{\rm I}$ and thus go from 0 to $x_{\rm d}$
where \beq x_{\rm d}=2x_{\rm max} \eeq is a suitable choice. However,
the numerical results are very stable to reasonable variations of the
choice of $x_{\rm d}$. Thus we rewrite
\beqa               
I_{\rm c}& =& T \int _{0}^{x_{\rm d}} dx \sqrt{xT} \exp (-x)
\frac{E_{\gamma}}{(xT-E_{\rm r})^{2}+\Gamma_{\rm tot}^{2}/4} \nonumber \\
         &+& T \int _{x_{\rm d}}^{\infty} dx \sqrt{xT} \exp (-x)
\frac{E_{\gamma}}{(xT-E_{\rm r})^{2}+\Gamma_{\rm tot}^{2}/4}
\eeqa               
So that we can use Gauss-Laguerre integration on the latter integral
we substitute $y=x - x_{\rm d}$ and obtain
\beqa               
I_{\rm c}& =& T \int _{0}^{x_{\rm d}} dx \sqrt{xT} \exp (-x)
\frac{E_{\gamma}}{(xT-E_{\rm r})^{2}+\Gamma_{\rm tot}^{2}/4} \\ &
&\mbox{}+T \int _{0}^{\infty} dy \sqrt{(y+x_{\rm d})T} \exp (-x_{\rm
d})\exp(-y) \frac{E_{\gamma}} {((y+x_{\rm
d})T-E_{\rm r})^{2}+\Gamma_{\rm tot}^{2}/4} \nonumber
\eeqa               
The first integral can now be calculated using Gauss-Legendre
integration and the last integral can be calculated by Gauss-Laguerre
integration.  Thus we obtain for the numerical evaluation of
Eq.~\ref{okpa_1}    
\beq                
I_{\rm c} =  \sum_{i=1}^{N_{\rm G}}
\omega^{\rm G}_iF^{\rm G}(x^{\rm G}_i)
+\exp(-x_{\rm d})   
\sum_{i=1}^{N_{\rm L}}\omega^{\rm L}_iF^{\rm L}(y^{\rm L}_i)
\label{sumnumer}    
\eeq                
where the superscripts G and L indicate Gauss-Legendre and
Gauss-Laguerre integration, respectively, and where
\beq                
F^{\rm G}(x) = T \sqrt{xT} \exp (-x)
\frac{E_{\gamma}}{(xT-E_{\rm r})^{2}+\Gamma_{\rm tot}^{2}/4} \nonumber
\label{gsum}        
\eeq                
and                 
\beq                
F^{\rm L}(y) = T \sqrt{(y+x_{\rm d})T}
\frac{E_{\gamma}} {((y+x_{\rm d})T-E_{\rm r})^{2}+\Gamma_{\rm tot}^{2}/4}
\label{lsum}        
\eeq                
Note that the exponential term is retained in $F^{\rm G}$ but not
in $F^{\rm L}$.     
\newpage            
 
\section{Details of the fluence determination in the Oklo reactor zones}
 
In this section the following notation is used:
\newcounter{bean}   
\begin{list}        
{\Roman{bean}}{\usecounter{bean}
\setlength{\leftmargin}{1.0in}
\setlength{\rightmargin}{0.0in}
\setlength{\labelwidth}{0.75in}
\setlength{\labelsep}{0.25in}
}                   
\item[$N_{143}$ ]   
The number of $^{143}$Nd atoms per unit volume after the Oklo reactor shutdown. In the
absence of contamination, in- and/or outflow of material this number is unchanged since
the shutdown.       
\item[$N_{147}$ ]   
The number of $^{147}$Sm atoms per unit volume after the Oklo reactor shutdown. In the
absence of contamination, in- and/or outflow of material this number is unchanged since
the shutdown.       
\item[$N_{235}$]    
The current number of $^{235}$U  atoms per unit volume in an  Oklo reactor zone.
\item[$N_{238}$]    
The current number of $^{238}$U  atoms per unit volume in an  Oklo reactor zone.
\item[$N^0_{235}$]  
The  number of $^{235}$U  atoms per unit volume in an  Oklo reactor zone at the start of
the reactor about two billion years ago. This number can simply be calculated
from data  observed now
at other uranium deposits where no nuclear reaction has occurred.
\item[$N^0_{238}$]  
The  number of $^{238}$U  atoms per unit volume in an  Oklo reactor zone at the start of
the reactor about two billion years ago. This number can simply be calculated
from data  observed now.
at other uranium deposits where no nuclear reaction has occurred.
\item[$I_A$] Resonance integral.
\item[$r_{\rm rz}$] Epi-thermal index. Accounts for the contribution of the neutron
spectrum integrated over low-lying resonances. The subscript rz indicates that
depends on the shape of the neutron spectrum and other
factors and is therefore different in each reactor zone (rz) at Oklo.
\item[$Y_{143}$] Cumulative fission yield of $^{143}$Nd.
\item[$Y_{147}$] Cumulative fission yield of $^{147}$Sm.
\item[$\hat{\sigma}_{\rm f235}$] $^{235}$U fission cross section.
\item[$\hat{\sigma}_{143}$] $^{143}$Nd total neutron-absorption cross section.
\item[$\hat{\sigma}_{147}$] $^{147}$Sm total neutron-absorption cross section.
\item[$\hat{\sigma}_{235}$] $^{235}$U total neutron-absorption cross section.
\item[$\hat{\sigma}_{238}$] $^{238}$U total neutron-absorption cross section.
\item[$\sigma^0_{A}$] Thermal neutron-capture cross section.
\item[$\lambda_{235}$] $^{235}$U decay constant.
\item[$\lambda_{238}$] $^{238}$U decay constant.
\item[$\hat{\phi}_{\rm rz}$] Average neutron flux in an Oklo reactor zone.
\item[$C_{\rm rz}$] Restitution factor of $^{235}$U in an Oklo reactor zone.
It accounts for the restitution of
$^{235}$U through neutron capture on $^{238}$U,
$\beta$-decay of $^{239}$U to $^{239}$Np,
$\beta$-decay of $^{239}$Np to $^{239}$Pu,
$\alpha$-decay of  $^{239}$Pu to $^{235}$U,
and smaller contributions from other reactions.
\item[$\Delta t_{\rm rz}$] Time duration of the neutron flux in the Oklo reactor.
\item[$t$] Time elapsed since the start of the Oklo reactor, about 2 billion years.
\end{list}          
 
The fluence $\hat{\phi}_{\rm rz} \Delta t_{\rm rz}$ in the different
Oklo reactor zones may be determined from
studies of the remnants of the reactions:
\beqa               
^{143}{\rm Nd}({\rm n},\gamma)^{144}{\rm Nd} \nonumber \\[1ex]
^{147}{\rm Sm}({\rm n},\gamma)^{148}{\rm Sm}  \nonumber\\[1ex]
^{235}{\rm U}({\rm n}, x)X  \;\; {\rm and} \nonumber \\[1ex]
^{238}{\rm U}({\rm n}, x)X \nonumber
\eeqa               
 
The absorption cross sections $\hat{\sigma}_{143}$ and
$\hat{\sigma}_{147}$ associated with the first two reactions may be expressed in the form
\beq                
\hat{\sigma}_A = \sigma_A^0 + r_{\rm rz}I_{A} \nnb
\eeq                
Specifically [29]   
\beqa               
\hat{\sigma}_{143} = 335 - 100r_{\rm rz} \nonumber \\[1ex]
\hat{\sigma}_{147} = \phantom{3}52 + 600r_{\rm rz} \nonumber
\eeqa               
in units of barn.   
The differential equations for the time-dependence of the number of atoms per unit volume $N_A$
of the isotopes involved in these reactions are
\beqa               
\frac{dN_{143}(t)}{dt}&=&-\hat{\sigma}_{143}\hat{\phi}_{\rm rz}N_{143}(t) +
N_{235}(t)Y_{143}\hat{\sigma}_{\rm f235}\hat{\phi}_{\rm rz} \nnb\\
\frac{dN_{147}(t)}{dt}&=&-\hat{\sigma}_{147}\hat{\phi}_{\rm rz}N_{147}(t) +
N_{235}(t)Y_{147}\hat{\sigma}_{\rm f235}\hat{\phi}_{\rm rz}\nnb\\
\frac{dN_{235}(t)}{dt}&=&-\lambda_{235}N_{235}(t) +
N_{235}(t)(1-C_{\rm rz})\hat{\sigma}_{\rm 235}\hat{\phi}_{\rm rz} \nnb\\
\frac{dN_{238}(t)}{dt}&=&-\lambda_{238}N_{238}(t) +
N_{238}(t)\hat{\sigma}_{\rm 238}\hat{\phi}_{\rm rz} \nonumber
\eeqa               
 
We obtain as solutions
\beqa               
N_{143} = \frac{N^{0}_{235} Y_{143} \hat{\sigma}_{\rm f235}}{\hat{\sigma}_{143}
-                   
(1-C_{\rm rz})\hat{\sigma}_{235}}
\left\{\exp\left[-(1-C_{\rm rz})\hat{\sigma}_{235} \hat{\phi}_{\rm rz} \Delta t_{\rm rz}\right]-
       \exp\left(-\hat{\sigma}_{143}\hat{\phi}_{\rm rz} \Delta t_{\rm
rz}\right)\right\}  
 \label{conc143} \\[1ex]
N_{147} = \frac{N^{0}_{235} Y_{147}\hat{\sigma}_{\rm f235}}{\hat{\sigma}_{147} -
 (1-C_{\rm rz})\hat{\sigma}_{235}}
\left\{\exp\left[-(1-C_{\rm rz})\hat{\sigma}_{235}\hat{\phi}_{\rm rz} \Delta t_{\rm rz}\right]-
        \exp\left(-\hat{\sigma}_{147}\hat{\phi}_{\rm rz} \Delta t_{\rm
rz}\right)\right\}  
 \label{conc47} \\[1ex]
\left( \frac{N_{235}}{N_{238}} \right) =
\left( \frac{N^0_{235}}{N^0_{238}} \right)
\exp\left[-(\lambda_{235}-\lambda_{238})t\right]
\exp\left\{-\left[(1-C_{\rm
rz})\hat{\sigma}_{235}-\hat{\sigma}_{238}\right]\hat{\phi}_{\rm rz}
 \Delta t_{\rm rz}\right\}
\eeqa               
 
One may easily obtain the ratio $N^0_{235} / N^0_{238}$ of $^{235}$U to $^{238}$U at Oklo
by observing that today the ratio in uranium mines is 0.007252 and use
\beq                
\left( \frac{N^0_{235}}{N^0_{238}} \right)
\left[\frac{\exp\left(-\lambda_{235}t\right)}
           {\exp\left(-\lambda_{238}t\right)}\right] = 0.007252 \nonumber
\eeq                
This leads to       
\beq                
\left( \frac{N_{235}}{N_{238}} \right) = 0.007252
\exp\left\{-\left[(1-C_{\rm
rz})\hat{\sigma}_{235}-\hat{\sigma}_{238}\right]\hat{\phi}_{\rm rz}
 \Delta t_{\rm rz}\right\}
\label{rattoday}    
\eeq                
for the  ratio $^{235}$U to $^{238}$U observed at Oklo today.
 
For each reactor zone at Oklo there are three unknowns that have to be determined:
the restitution factor $C_{\rm rz}$, the epi-thermal index $r_{\rm rz}$, and the
neutron fluence $\hat{\phi}_{\rm rz} \Delta t_{\rm rz}$.
They can now be obtained as solutions to the three coupled
equations (\ref{conc143}), (\ref{conc47}), and
(\ref{rattoday}).   
Conditions for the occurrence of
 fission processes are influenced by many factors, such as the U
 content, neutron-absorber elements, water content among others. This
 explains why we often find different neutron fluences in different
 reactors.          
\newpage            
\mbox{ } \\         
\mbox{ } \vspace{-0.5in} \mbox{ } \\
 \begin{center}                                                                 
 {\bf References}                                                               
 \end{center}                                                                   
 \newcounter{bona}                                                              
 \begin{list}%
 {\arabic{bona})}{\usecounter{bona}                                             
 \setlength{\leftmargin}{0.5in}                                                 
 \setlength{\rightmargin}{0.0in}                                                
 \setlength{\labelwidth}{0.3in}                                                 
 \setlength{\labelsep}{0.15in}                                                  
 \setlength{\itemsep}{-0.03in}                                                  
 }                                                                              
 \item                                                                          
Y. Fujii, {\sl Oklo Phenomenon Revisited --- Strong Interaction Coupling        
  Constant Time-Dependent?} in Proceedings of International Workshop on         
  Gravitation and Astrophysics, ICRR, University of Tokyo, 17-19 November,      
  1997.                                                                         
                                                                                
 \item                                                                          
P.A.M. Dirac, Proc. Roy. Soc. {\bf A165} (1938) 199.                            
                                                                                
 \item                                                                          
F. Hoyle, {\sl Galaxies, Nuclei and Quasars}, Heinemann (London), 1965.         
                                                                                
 \item                                                                          
F.J. Dyson, Phys. Rev. Lett. {\bf 19} (1967) 1291.                              
                                                                                
 \item                                                                          
P.C.W. Davies, J. Phys. {\bf A5} (1972) 1296.                                   
                                                                                
 \item                                                                          
A.I. Shlyakhter, Nature {\bf 264} (1976) 340.                                   
                                                                                
 \item                                                                          
A.I. Shlyakhter, ATOMKI Report A/1 (1983), unpublished.                         
                                                                                
 \item                                                                          
L.L. Cowie and A. Songalia, Ap. J. {\bf 453} (1995) 596.                        
                                                                                
 \item                                                                          
A. Godone, et al, Phys. Rev. Lett. {\bf 71} (1993) 2364.                        
                                                                                
 \item                                                                          
R.W. Hellings, et al, Phys. Rev. Lett. {\bf 51} (1983) 1609.                    
                                                                                
 \item                                                                          
See, for example, The Oklo Phenomenon, Proc. of a Symposium, Libreville, June,  
  1975 (IAEA, Vienna, 1975).                                                    
                                                                                
 \item                                                                          
Natural Fission Reactors, Proc. of a meeting on natural fission reactor, Paris, 
  France, December, 1977 (IAEA, Vienna, 1978).                                  
                                                                                
 \item                                                                          
T. Damour and F. Dyson, Nucl. Phys. {\bf B480} (1996) 37.                       
                                                                                
 \item                                                                          
Neutron Cross Sections, Volume I (Academic Press, New York, 1984).              
                                                                                
 \item                                                                          
M. Lucas, R. Hagemann, R. Naudet, C. Renson and C. Chevalier,Natural Fission    
  Reactors, Proc. of a meeting on natural fission reactor, Paris, France,       
  December, 1977 (IAEA, Vienna, 1978) p. 407.                                   
                                                                                
 \item                                                                          
JENDL Nuclear data library, URL http://wwwndc.tokai.jaer.go.jp/NuC/index.html.  
                                                                                
 \item                                                                          
H. Hidaka and P. Holliger, Geochim. Cosmochim. Acta {\bf 62} (1998), 89.        
                                                                                
 \item                                                                          
D. Curtis, T. Benjamin, A. Gancarz, R. Loss, K. Rosman, J. DeLaeter, J.E.       
  Delmore and W.J. Maeck, Applied Geochemistry, {\bf 4} (1989) 49.              
                                                                                
 \item                                                                          
H. Hidaka and F. Gauthier-Lafaye, unpublished data (1998).                      
                                                                                
 \item                                                                          
P. Holliger and C. Deviller, Earth Planet. Sci. Lett. {\bf 52}(1981),
76.

 \item
F. Gauthier-Lafaye, F. Weber and H. Ohmoto, Econ. Geol. {\bf 84} (1989) 2286.   
 \item
L. Pourcelot and F. Gauthier-Lafaye, Chem. Geol. {\bf 157]} (1999), 155.

 \item
S. Perlmutter, G. Aldering, G. Goldhaber, R.A. Knop, P. Nugent, P.G. Castro, S. 
  Deustua, S. Fabbro, A. Goobar, D.E. Groom, I. M. Hook, A.G. Kim, M.Y. Kim,    
  J.C. Lee, N.J. Nunes, R. Pain, C.R. Pennypacker, R. Quimby, C. Lidman, R.S.   
  Ellis, M. Irwin, R.G. McMahon, P. Ruiz-Lapuente, N. Walton, B. Schaefer, B.J. 
  Boyle, A.V. Filippenko, T. Matheson, A.S. Fruchter, N. Panagia, H.J.M.        
  Newberg, W.J. Couch, ApJ, to be published.                                    
                                                                                
 \item
Y. Fujii and T. Nishioka, Phys. Rev. {\bf D42} (1990) 361.                      
                                                                                
 \item
Y. Fujii, M. Omote and T. Nishioka, Prog. Theor. Phys. {\bf 92} (1994), 521.    
                                                                                
 \item
Y. Fujii, Astrop. Phys. {\bf 5} (1996) 133.                                     
                                                                                
 \item
Y. Fujii, {\sl $\Omega_{\Lambda}$ and a new type of dissipative structure},     
  delivered at XXXIIIrd Rencontres de Moriond, Fundamental Parameters in        
  Cosmology (Les Arcs, France, January 17-24, 1998) and papers cited therein.   
                                                                                
 \item 
J.K. Webb, V.V. Flambaum, C.W. Churchill, M.J. Drinkwater and J.D. Barrow,      
  Phys. Rev. Lett. {\bf 82} (1999) 884.                                         
                                                                                
 \item
J.D. Barrow and F.J. Tipler,{\sl The Anthropic Cosmological Principle},         
  Clarendon Press, Oxford, 1986.                                                
                                                                                
 \item
C.H.\ Westcott, Effective cross-section values for well-moderated thermal       
  reactor spectra, CRRP-960 (1960) (A.E.C.L., Chalk River).                     
                                                                                
 \end{list}                                                                     

\end{document}